\documentclass[10pt, journal,compsoc]{IEEEtran}
\usepackage{amsmath,amsfonts}
\usepackage{algorithmic}
\usepackage{algorithm}
\usepackage{array}
\usepackage[caption=false,font=normalsize,labelfont=sf,textfont=sf]{subfig}
\usepackage{textcomp}
\usepackage{stfloats}
\usepackage{url}
\usepackage{verbatim}
\usepackage{graphicx}
\usepackage{cite}
\usepackage{enumitem}
\usepackage{multirow}
\usepackage{tabularx}
\usepackage{colortbl}
\usepackage[table,xcdraw]{xcolor}

\makeatletter
\def\Cline#1#2{\@Cline#1#2\@nil}
\def\@Cline#1-#2#3\@nil{%
  \omit
  \@multicnt#1%
  \advance\@multispan\m@ne
  \ifnum\@multicnt=\@ne\@firstofone{&\omit}\fi
  \@multicnt#2%
  \advance\@multicnt-#1%
  \advance\@multispan\@ne
  \leaders\hrule\@height#3\hfill
  \cr}
\makeatother

\newcommand{\HDY}[1] 
{\textcolor{black}{#1}}

\usepackage{tabularx}

\begin{document}

\title{Evaluating Effectiveness of Interactivity in Contour-based Geospatial Visualization\HDY{s}
}

\author{
Abdullah-Al-Raihan Nayeem$^\dag$,  Dongyun Han$^\dag$, William J. Tolone, and Isaac Cho$^*$, \textit{IEEE member}
\IEEEcompsocitemizethanks{\IEEEcompsocthanksitem 
 Dongyun Han and Isaac Cho are with Utah State University E-mail:{{dongyun.han}@usu.edu, {isaac.cho}@usu.edu} \protect\\
 \IEEEcompsocthanksitem 
 Abdullah-Al-Raihan Nayeem is with Corteva Agriscience
 {{raihan.nayeem}@gmail.com} \protect\\
\IEEEcompsocthanksitem William J. Tolone is with the University of North Carolina at Charlotte. E-mail:{{william.tolone}@charlotte.edu} \protect\\
\IEEEcompsocthanksitem $^\dag$ These authors contributed equally, and the order of authors does not indicate the extent of their contributions. \\ 
\IEEEcompsocthanksitem* corresponding author}
\thanks{Manuscript received April 19, 2021; revised August 16, 2021.}}

\markboth{Journal of \LaTeX\ Class Files,~Vol.~14, No.~8, August~2021}%
{Shell \MakeLowercase{\textit{et al.}}: A Sample Article Using IEEEtran.cls for IEEE Journals}

\IEEEpubid{0000--0000/00\$00.00~\copyright~2021 IEEE}

\IEEEtitleabstractindextext{%
\begin{abstract}
Contour maps are an essential tool for exploring spatial features of the terrain, such as distance, directions, and surface gradient among the contour areas. 
User interactions in contour-based visualizations create approaches to visual analysis that are noticeably different from the perspective of human cognition. 
As such, various interactive approaches have been introduced to improve system usability and enhance human cognition for complex and large-scale spatial data exploration.
However, what user interaction means for contour maps, its purpose, when to leverage, and design primitives have yet to be investigated in the context of analysis tasks.
Therefore, further research is needed to better understand and quantify the potentials and benefits offered by user interactions in contour-based geospatial visualizations designed to support analytical tasks. 
In this paper, we present a contour-based interactive geospatial visualization designed for analytical tasks. 
We conducted a crowd-sourced user study (N=62) to examine the impact of interactive features on analysis using contour-based geospatial visualizations. 
Our results show that the interactive features aid in their data analysis and understanding in terms of spatial data extent, map layout, task complexity, and user expertise. 
Finally, we discuss our findings in-depth, which will serve as guidelines for future design and implementation of interactive features in support of case-specific analytical tasks on contour-based geospatial views.
\end{abstract}

\begin{IEEEkeywords}
Contour Visualization, User Interaction for Visualization Analytics, Crowd-Sourced User Study
\end{IEEEkeywords}}

\maketitle

\section{Introduction}
\maketitle
\label{sec_introduction}
\IEEEPARstart{G}{eospatial} 
visualizations have become indispensable tools for supporting the analysis of spatial variables across a wide range of research domains, including atmospheric science \cite{McLean2020, Alder2015, li2020sovas}, meteorology \cite{wang2014, Sharma2018, wang2019open}, and urban planning \cite{8613864, zeng2014visualizing, feng2020topology}. 
Contour maps, a traditional cornerstone of geospatial visualizations, enable a complex analysis of geographical events (hereafter referred to as ``events''). 
These events include atmospheric and meteorological phenomena such as landform shapes, mountain elevations, ocean depths, and climate projections. 
In recent years, collaborations between visualization researchers and domain experts have intensified, leading to multidisciplinary research.
These collaborative efforts have been instrumental in developing effective geovisualizations and interaction techniques to identify hidden patterns \cite{Steed2020, nayeem2021visual, maceachren1997exploratory, andrienko2007geo, andrienko2007geo, robinson2017}. However, compared to other forms of geovisualizations such as Choropleth \cite{andrienko2001choropleth} and Cartogram \cite{nusrat2016state}, the exploration and interaction potential of contour maps have not been fully realized.

Earlier research \cite{tominski2011information, afzal2019state, CHEN2019129, andrienko2020spatio, nayeem2022dcpviz, gdlviz2022} indicates that many domain experts relied on traditional static depictions of geospatial data in contour visualizations \cite{gibson2017use, Lee2014}. 
Recently, improved spatial resolution in climate projection models has led to better representations of extreme weather events \cite{Lee2014}, the hydrological cycle \cite{Lee2017}, and important land surface processes \cite{DeSales2013}.
Although contour maps can present 3D information in 2D and provide a topographic view, they have limitations when dealing with finer spatial resolution data \cite{nayeem2022dcpviz}. Such contour maps can become cluttered and often require a certain level of expertise to understand and gain insights from finer spatial resolution.

In this context, user interactivity with contour maps has the potential to enhance the efficacy of such analyses. Furthermore, it is known that presenting more views (maps in this study) is beneficial for users in their sensemaking~\cite{andrews2010space, bradel2013large} and knowledge development~\cite{leigh2019usage} by providing a larger volume of information.  
For instance, they could assist users in identifying specific events or outlines and illustrating the relationships between their features based on user interactions. 
Nonetheless, quantifying these advantages in \HDY{analyses} presents a significant challenge. User interactions and visual representations could play a crucial role in organizing, structuring, and segmenting spatial data, facilitating user exploration from abstract overviews to focused inspections. 
They manifest in geospatial visualizations introducing new approaches to visual analysis, distinctively impacting human cognition.

To evaluate the usability of interactivity in contour maps, we formulated the following research questions (RQs):
\begin{enumerate}[label=RQ\arabic*., leftmargin=* ]
    \item What advantages in performance are expected when utilizing interactive contour maps in contrast to traditional static contour maps?
    
        \item Which interactive features are the most useful or frequently used \HDY{in analytical tasks}?

\end{enumerate}

To address these research questions, we conducted a crowd-sourced user study to evaluate the interactive features of contour maps. 
Drawing upon a literature review on previous research \cite{OGAO200223, koua2006,nayeem2021visual, nayeem2022dcpviz, gdlviz2022}, we selected seven tasks and implemented a web-based user interface utilizing two real-world geospatial datasets for the study. 
Our findings suggest that the participants achieved greater accuracy using the interactive map compared to the static contour map, particularly in tasks such as associating, categorizing, clustering, identifying, and ranking the geospatial data points. 
Moreover, the results reveal that inspection and selection are the most frequently used interactive functions, followed by annotation, filter/search, and encode. 
These features are tailored to specific task types, such as annotations being mostly used in \textit{Locate, Rank,} and \textit{Associate} tasks, while filter interactions are frequently utilized in \textit{Identify} and \textit{Categorize} tasks. 
The results also suggest that these features benefit novice users, closing their performance gaps in complex \HDY{tasks.} 
These findings can be utilized as design considerations for crafting and assessing interactive contour maps designed for particular use cases related to \HDY{geospatial analysis}.

\begin{figure*}[t]
    \centering
    \includegraphics[width=0.9\linewidth]{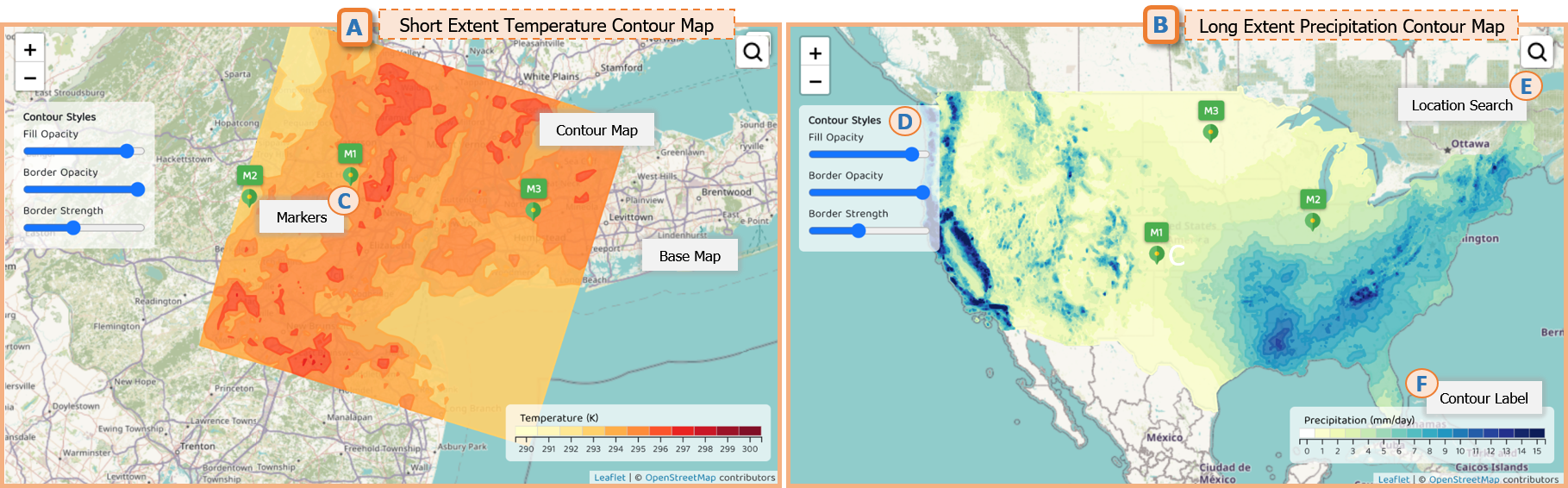}
    \caption{
    Illustration of two interactive contour-based geospatial visualizations:  (A) temperature intensity in New York City and (B) precipitation intensity in the contiguous U.S. recorded at a certain time. 
    The interactive features are intended to support \HDY{visual analysis} of the contour regions.
    }
    \label{fig:map-view}
\end{figure*}

\section{Related Works}

\subsection{Geospatial Visualization}

Visualization researchers have introduced many geovisual environments for various data types, analysis requirements, and use cases. 
The potential of geospatial data for mapping and analytics, enabled by geographic information systems, has been realized in fields beyond earth science.
Such domains, including traffic systems, energy systems, health informatics, and entertainment,  have heavily relied on location or geographical data \cite{poweroflocation}. Choropleth, point map, cluster map, cartogram maps, etc. are popularly employed to visualize the data with spatial dimensions \cite{Maskey2020, Tominski2021, nayeem2022dcpviz} depending on analytic necessity. 
While choropleth maps are a common way to display discrete interval or ratio scaled data \cite{gdlviz2022}, heatmap and distribution maps often illustrate continuous data variables containing geo-coordinated attributes  \cite{Quinan2016}. Despite this variety, to the best of our knowledge, we have yet to encounter any interactive contour map specifically designed to provide a \HDY{geovisual environment} for spatial data analysis.

Traditionally, users have relied on static geospatial depictions in which information is visually estimated from traditional geospatial views, typically in the form of raster images. 
Notable examples of tools facilitating this approach included NASA's Panoply software package, which extracts high-dimensional climate datasets from memory-efficient encodings and visualizes them as geospatial contours \cite{2018AGUFMIN21B35A}, and  QGIS \cite{Hugentobler2008}, a widely used open-source cross-platform software that visualizes geospatial data in various format (e.g., GeoJSON, NetCDF, Shapefile) as a mesh layer. In addition, toolkits such as MATLAB \cite{hanselman1996mastering}, NCAR \cite{NCAR2019}, and GrADS \cite{Berman2001} remain popular for visualizing geospatial data. 
Geospatial contours generated by these toolkits are frequently used by domain researchers to illustrate the results of their analyses \cite{gibson2017use, CAO2021145320}. However, these traditional approaches are often static, limiting the scope and depth of data interactions and exploration.

Moreover, we have examined the transformations of geospatial visualizations to accommodate multi-dimensional attributes. For instance, Community Data Analysis Tools (CDAT) provide a software infrastructure and platform for analyzing multidimensional data \cite{williams2002climate, potter2009visualization}, which is widely used for distributed analyses of geospatial data. In addition, CDAT offers a visualization and control system that interacts with Python to visualize geospatial data \cite{williams2002climate}. GIS4WRF \cite{MEYER2019166} is another toolkit for processing and visualizing weather data and forecasting NetCDF in an interactive single graphical environment. MeteoInfo is a desktop application that provides a geospatial visualization platform for multidimensional meteorological data \cite{wang2014}. Moreover, ClimateData.US provides an interactive geovisual interface that provides comparative visualization of climate data in different emission scenarios \cite{Herring2017}. Geospatial views are developed for visualizing time-varying data and performing rule-based associations \cite{wang2018}. 
However, these visualization approaches often lack adequate interactive features for diverse explorations of spatial data variables.

\subsection{Evaluating Geovisual Environments}

The evaluation criteria of geovisual environments range from the ability to display the data, the interactivity for analytical tasks, and the self-explanatory attributes of the visualizations \cite{johansson2010evaluating}. It generally involves visualization researchers developing geovisual components followed by user experts interacting with it and providing constructive feedback for the next iteration or outlining the usability for intended analysis purposes \cite{nayeem2021visual, nayeem2022dcpviz, gdlviz2022}. 
For example, Koua et al. \cite{koua2006} conducted a user study with geographers, cartographers, geologists, and environmental scientists and evaluated the usability of different visualization methods for geovisual analyses. 
Their study included visualization such as parallel coordinates, and self-organizing heatmap representation against the geospatial map to evaluate users' performance in \HDY{10 analytical tasks} such as categorizing, clustering, and identifying spatial areas.

In our review, we found a small amount of literature that quantitatively evaluated the interactivity of geovisual environments with users from a diverse range of expertise and the effect on their performance.
Nagel et al. \cite{nagel2012} introduced a tabletop map visualization that enables users to explore geospatial networks interactively. Mahmood et al. \cite{mahmood2019improving} presented a mixed reality geospatial visualization method that fosters remote collaboration and knowledge sharing for sensemaking tasks. Herman et al.~\cite{herman2018evaluation} evaluated the interactivity of 3D maps for tasks such as landscape management, and crisis management. They suggested that interactive 3D maps are better suited for complex tasks, allowing users to make more accurate decisions in less time. 
Amini et al.~\cite{amini2014impact} pointed out that 2D maps have limitations in visualizing movement data, such as car and plane trajectories when trajectories overlap. They investigated the effect of interactive 2D and 3D maps on analyzing movement data and reported that participants performed better with the 3D map than with the 2D map. \HDY{Lateh and Raman~\cite{lateh2005study}} examined the static and interactive 2D maps in learning and teaching geography in a high school. They found that the interactive map significantly enhances the students' ability to read and understand spatial information on the map. However, earlier studies investigated the effect of map interactivity on limited tasks such as value identification and finding clusters. 

Roth \cite{roth2013empirically} developed a functional taxonomy of interaction for cartography and geospatial visualizations. The preemptive definitions from the study were separated into interaction objectives and operators, which serve as a structure to study user interactions for map-based visualizations. The foundational knowledge for designing cartographic view outlined in Roth \cite{roth2013empirically} and \HDY{the geospatial task} classification presented by Koua et al. \cite{koua2006} are relevant to the scope of this study design for quantitative evaluation.

\section{Contour-based Geovisualization}

This section defines terms and describes specific tasks and interactive features for contour-based geovisualization. 

\subsection{Interactive and Static Contour Map}
The geospatial map is a multi-layered view, comprising a base map, contour map, and spatial masks. 
The base map signifies a collection of geographic features that illustrate the background scene for a given location. 
The contour map provides a topographic representation of a location by clustering areas of identical value with lines and polygons \cite{hahmann2015contour}. Next, the spatial masks outline boundaries for the contiguous US regions, providing a visual segmentation of area-specific (regional) intensity.

Within the contour layer, boundaries create multiple polygons that we identify as contour regions. For example, the contour maps in Fig. \ref{fig:map-view}A and B illustrate the temperature intensity for New York City, and the precipitation intensity for the contiguous US respectively. On one hand, we define an interactive contour map as a map that has the capability to include interactive features. 
An interactive contour map, for example, could have the capability to present the total area of regions corresponding to unique intensity values. 
This allows the user to measure the intensity distribution or spread within the area of interest based on selected values. 
On the other hand, we \HDY{define} a static contour map as one that does not offer the capacity to accommodate interactive features beyond the traditional approach.

\subsection{\HDY{Visualization Analysis Tasks}}\label{sec:visual-tasks}

We reviewed a set of \HDY{analysis tasks} previously used in evaluating \HDY{geovisual environments} \cite{keller1994visual, koua2006}. 
This study utilizes the exhaustive list proposed by Keller et al. \cite{keller1994visual} and subsequently employed by Koua et al. \cite{koua2006} for a usability study in \HDY{contour-based geovisualization.} 
The identified tasks are low-level analytical tasks that are common among climate and environmental scientists \cite{koua2006, nayeem2022dcpviz, zahan2021contour, gdlviz2022}. 
Although these tasks are applicable to various types of map visualizations, we selected the following seven tasks because they are suitable for the contour map based on their operations:

\begin{enumerate}[leftmargin=* ]
    \item \textit{\textbf{Associate}} tasks involve users in finding similarities between data points based on the characteristics of attributes, geographic locations, and patterns of events.

    \item \textit{\textbf{Categorize}} tasks require defining all regions on the display by the boundaries. The regions are usually categorized by common features or spatial positioning.

    \item \textit{\textbf{Cluster}}-type tasks require users to identify different groups of similar data points or regions within the geospatial data. 

    \item \textit{\textbf{Distinguish}} analytical tasks require users to identify the differences or variations in geospatial areas or locations.

    \item \textit{\textbf{Identify}} demands establishing the relationship between data attributes based on their shared characteristics.

    \item \textit{\textbf{Locate}} is identifying a certain range of data points or events on the map. 

    \item \textit{\textbf{Rank}}-type tasks require feature ordering based on data attributes, clusters, or geographic locations.
\end{enumerate}

\subsection{Interactive Features} \label{sec:user-interactions}

The user interactions are designed to support \HDY{analysis tasks} by facilitating human cognition processing \cite{lu2017state}.
To find these interactive features, we gathered insights from literature review \cite{nayeem2022dcpviz, gdlviz2022, nayeem2021visual, wang2018, Herring2017, McLean2020}. 
For example, DiBiase's Swoopy schematic \cite{dibiase1990visualization} provided a framework, that includes exploration, analysis, synthesis, and presentation, to identify the purpose of maps. 
Based on previous research, we identified a set of interactive features including \textit{\textbf{Inspection, Selection, Annotation, Filter, Encoding,}} and \textit{\textbf{Synchronization}}. In the following section, we discuss them in detail along with our implementation.

Please note that interactive maps usually provide navigational features like zooming and panning. However, we excluded them in this work because our major concern is to investigate the usability of user interactions with contour regions. 
\section{Web Interface for user study }

\begin{figure}[t]
  \centering
  \includegraphics[width=1\linewidth]{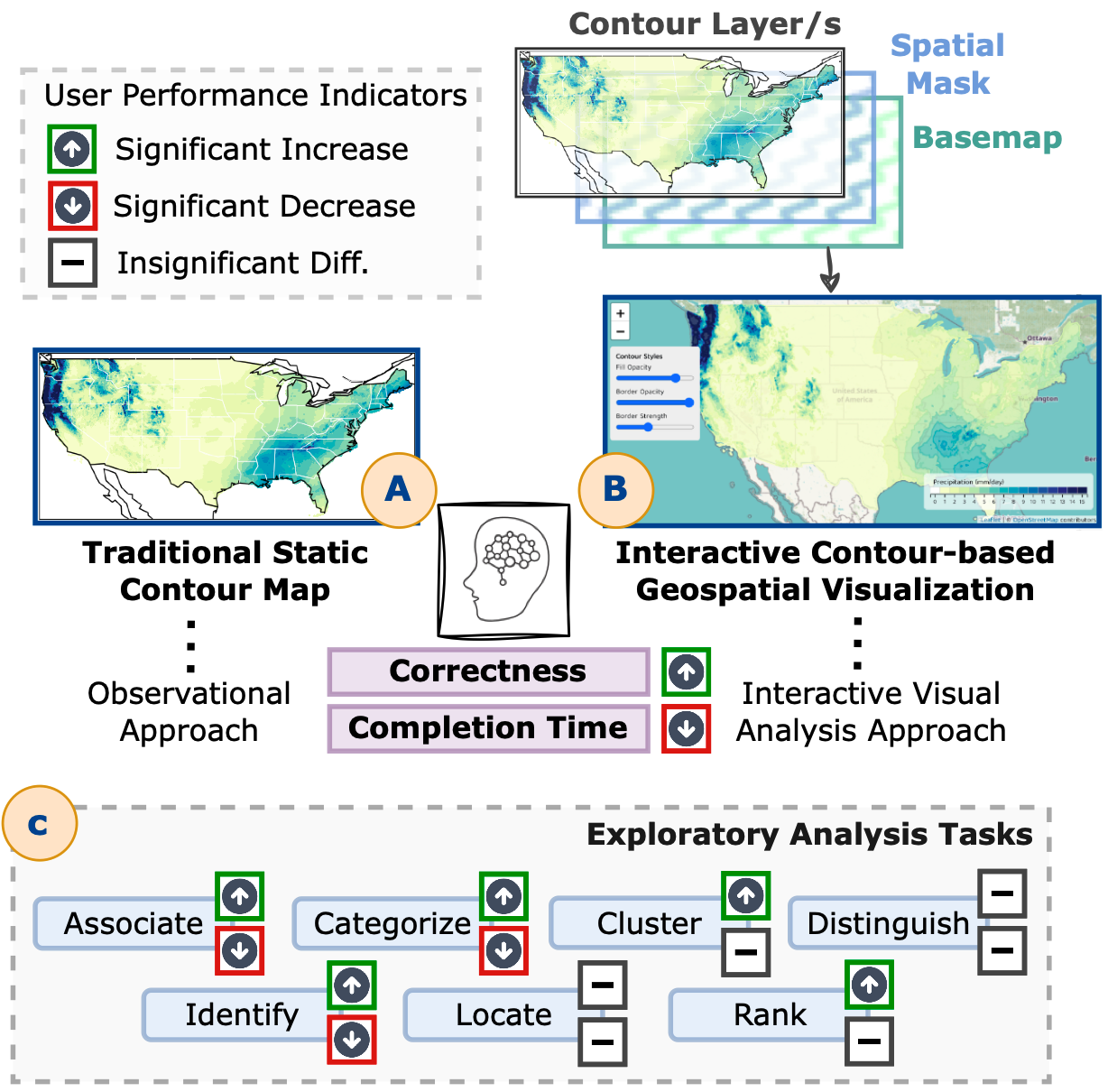}
  \caption{
    Overview of the quantitative evaluation for contour-based interactive geospatial visualization: We conducted a crowd-sourced study to measure user performance \HDY{in analysis} tasks. 
    We compared (A) a traditional static contour map with (B) a contour-based interactive geospatial visualization. The results (C) suggest that interactive features significantly improve performance in complex tasks such as spatial data association, categorization, clustering, and ranking.
  }
  \label{fig:interaction-teaser}
\end{figure}

For our user study, we designed and developed a web interface that included multiple coordinated visual and interactive components (Fig. \ref{fig:interaction-teaser}) \footnote{ \url{https://vizus.cs.usu.edu/app/geospatial-study}}. 
The interface is built with React.js and incorporates D3.js and Leaflet.js for interactive visualizations. 
It utilizes OpenStreetMap (OSM) \cite{openstreetmap} as its base map to show geographical features such as boundaries, rivers, and highways. 
It is worth noting that OSM employs the Web Mercator projection, which introduces considerable distortions in regions near the poles, potentially complicating area comparisons. 
However, since our maps are primarily focused on regions within the US, the impact of this limitation was substantially reduced. Django and MongoDB are utilized for the API server and database to store user interaction logs.

\subsection{Geospatial Datasets}
Two separate geospatial datasets are employed to visualize contour maps for distinct scenarios:

\begin{itemize}[leftmargin=* ]
    \item \textit{\textbf{NY Temperature:}} Hourly-averaged temperature forecast for the tri-state area (Fig. \ref{fig:map-view}A) based on the Weather and Research Forecasting (WRF) model published by Coastal Urban Environmental Research Group (CUREG) \cite{nytemperaturedata}
    
    \item \textit{\textbf{CONUS Precipitation:}} Monthly-averaged precipitation projections for the CONUS published by NASA Earth eXchange (NEX), where downscaled climate projections (NEX-DCP30) were generated in different emission scenarios such as RCP 2.8, RCP 3.5 and RCP 4.5 \cite{Nemani2015} (Fig. \ref{fig:map-view}B)
    
\end{itemize}

\subsection{Interactive Geospatial Map and Interactions}
The interface displays one or two contour-based geospatial maps based on a study condition discussed in Section~\ref{section_map_layout}. When multiple maps are provided, they are uniformly either static or interactive. 
When a contour-based map is interactable, it supports the interactions identified in Section~\ref{sec:user-interactions}. 
Please note that \HDY{the} \textit{\textbf{Synchronization}} feature is always supported when two interactive maps are presented in the interface.
The following demonstrates our implementation of the interactive features.

\begin{itemize}[leftmargin=* ]
\item \textit{\textbf{Inspection}} allows the user to investigate a contour region. When the user hovers the pointer over the region, it highlights all the associated contour regions on the map and shows a tooltip. 
The tooltip provides additional information about the interacted region, the number of contour regions occupied by that intensity, and the contour area distribution for those regions.

\item \textbf{Selection} or deselection of a region of interest is accomplished by clicking the mouse on the map or one of the contour labels (Fig. \ref{fig:map-view}F). The region of interest can be either a contour region or a location on the map.
Following the selection, the contour area and regions are recalculated to reflect the updated selection.

\item \textit{\textbf{Annotation}} places a marker displaying details such as intensity, occupied cluster area, and comparative measures. 
A right mouse click on a point of interest enables the user to place a marker on the map, as depicted in Fig. \ref{fig:map-view}C. 
The view allows multiple markers across different locations. When a user clicks one of the markers with the mouse's left button, it displays contour information for that marker as well as other markers to aid in comparison.

\item \textit{\textbf{Filter/Search}} aids the user in locating specific areas or boundaries on the map that might otherwise be difficult to find manually. 
The interactive map features a search icon positioned at its top right corner, as illustrated in Fig. \ref{fig:map-view}E. The user can type the name of the desired area, and select the appropriate option from the provided list. 
A marker will then be placed on the chosen location. 
The marker operates in the same manner as annotation markers.

\item \textit{\textbf{Encoding}} enables the user to control the visibility of the contour area or border. The interactive map provides three sliders to control fill opacity, border opacity, and border strength as shown in Fig. \ref{fig:map-view}D.
This improves the user's ability to better correlate an event with its geographical location, as well as, allows the user to change the visual configuration to perceive insights from a cluttered view.

\item \textit{\textbf{Synchronization}} makes it easier to compare two contour maps, allowing \HDY{for analysis} of potential differences at points of interest. In this configuration, all previously discussed interactions are synchronized, which means that any user interaction with one contour map has a complementary effect on another. 

\end{itemize}

\section {Crowd-sourced user study}

Our study design is influenced by a literature review on crowd-sourced studies evaluating visualization components \cite{heer2010, nagel2012, karduni2021} that provided a well-grounded idea on participant recruitment and exclusion criteria, training requirements, and study procedures.

\subsection{Participants \& Recruitment}
\begin{figure}[t]
    \centering

    \includegraphics[width=0.9\linewidth]{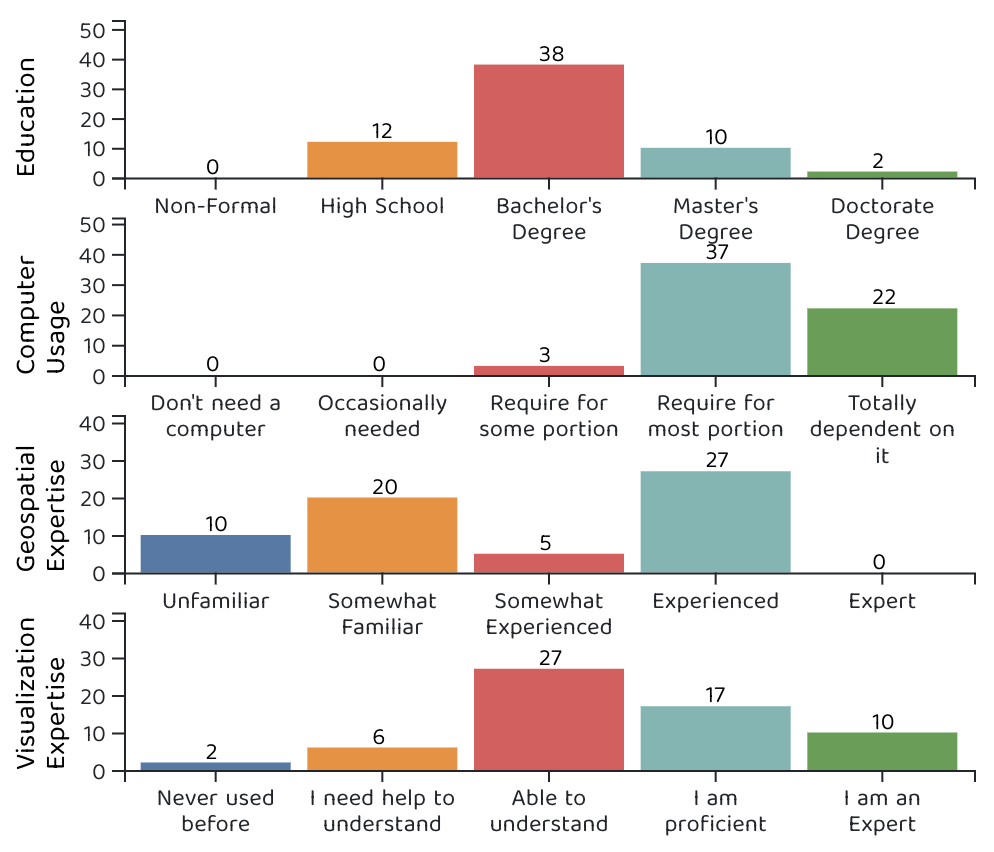}
    \caption{Summary of 62 participants' educational qualifications, computer usage, geospatial expertise, and visualization expertise. }
    \label{fig:demographic-data}
\end{figure}
We host the study online, with participants sourced from Amazon Mechanical Turk (MTurk). Our user study interface is directly linked to a Human Intelligent Task (HIT) we posted on the MTurk portal. 
To ensure the quality of the collected data, we selected participants from MTurk who had approval ratings above 95\% and were identified as master workers. \HDY{We initially recruited a total of 67 participants, all based in the United States.} In line with our pre-registration criteria, we exclude \HDY{four participants} who failed engagement questions and \HDY{another participant} who completed the study in under 8 minutes. This led to a final sample size of 62 participants (25 females and 37 males).
We ask the participants about their academic qualifications, daily computer usage, and how they rate their expertise in perceiving visualizations and analyzing geospatial data. 
For computer usage, visualization expertise, and geospatial analysis expertise, the participants identify themselves from the groups shown in Fig. \ref{fig:demographic-data}. 
They identified themselves as having various levels of expertise in visualization and geospatial analysis (Fig. \ref{fig:demographic-data}). 
Each participant earned a \$4 incentive for their contribution to our user study.
On average, participants took 8.6 minutes (\textit{SD=7.99}) for the training session and 29.14 minutes (\textit{SD=19.75}) to complete the main session.

\subsection{Study Factors}

\subsubsection{Map Type}

In this work, we divide map types into two categories: static map (Fig. \ref{fig:interaction-teaser}A) and interactive map (Fig. \ref{fig:interaction-teaser}B), based on their level of interactivity. The interactive map supports the aforementioned interactive features. 
Please keep in mind that the interactive map provides all the features, and we do not exclude or omit any specific features. We made this decision to ascertain and discuss which interactive features participants commonly utilize to accomplish the tasks. Conversely, the static map is designed considering the traditional approach of contour maps and only offers panning and zooming capabilities.

\subsubsection{Map Layout}\label{section_map_layout}
The map layout includes two conditions: single-map and multiple-map. 
The single-map condition provides only one static or interactive map, as the name implies. The multiple-map condition displays two maps side by side, with each map potentially containing data from different locations or time periods. We refer to them as Snapshot Left/Right. They are either static or interactive maps depending on the map types. For the interactive multiple-map, synchronized interactions are designed to help users compare events across different maps. It is commonly used in temporal-spatial data research. 


\bgroup
\def\arraystretch{1.2}%
\begin{table*}[h]
\centering
\caption{The participants performed \HDY{these geovisual analyses} from seven unique task types. \HDY{These tasks} were designed for both single and multiple-map layouts to evaluate the usability of interactive features in a geovisual environment.
}
\begin{tabular}{| p {5em}| p {24em} | p{25em} |}
\hline
\textbf{Task Type} & \textbf{single-map}                                                                                                                                                                                          & \textbf{multiple-map}                                                                                                                                                                            \\ \hline
Associate          & Based on the area covered which two \emph{[variable/marker]} intensities below are the most similar?                                                                                    & Find the \emph{[variable/marker]} intensities from both snapshot that cover the \emph{[largest/smallest]} area.                                        \\ \hline
Categorize         & What is the \emph{[area/intensity]} (difference) for the \emph{[highest intensity/largest area]} recorded on \emph{[contour/Location X]}? & What is the area difference of \emph{[variable]} \emph{[intensity]} between Snapshot Left and Right?                                                   \\ \hline
Cluster            & How many \emph{[unique/total]} contour regions are visible on the map?                                                                                                                  & How has the number of \emph{[unique/total]} contour regions changed from `Snapshot Left' to `Snapshot Right'?                                                               \\ \hline
Distinguish        & What is the \emph{[area/intensity]} difference between the highest and lowest \emph{[variable]}?                                                                   & How does the \emph{[highest/lowest]} intensity shift from `Snapshot Left' to `Snapshot Right'?                                                                              \\ \hline
Identify           & What is the \emph{[intensity/difference between]} \emph{[variable]} in Location \emph{[X]} (and Location \emph{[Y]})?    & What is the intensity difference between Location \emph{[X]} in `Snapshot Left' and Location \emph{[Y]} in `Snapshot Right'?                           \\ \hline
Locate             & What is the \emph{[highest/lowest]} intensity on the \emph{[contour/marked area]}?                                                                                 & What is the \emph{[highest/lowest]} intensity \emph{[between both snapshots/among all markers]}?                                                       \\ \hline
Rank               & Order the \emph{[markers/intensities below]} from \emph{[high to low/low to high]} \emph{[intensity/area]}.                                   & Order the snapshots based on the \emph{[larger/smaller]} area covered by the \emph{[lowest/highest]} \emph{[variable]} intensity. \\ \hline
\end{tabular}
\label{tab:exploratory-tasks}
\vspace{-.3cm}
\end{table*}

\subsubsection{\HDY{Visualization Analysis Tasks}}
In this work, we consider the seven \HDY{visualization tasks} identified in Section~\ref{sec:visual-tasks}. 
\HDY{These tasks} can be broadly categorized as simple or complex tasks, depending on the number of data points they encompass.
The simple tasks involve more direct spatial exploration, primarily focusing on Distinguish, Identify, and Locate task types. Here, users typically interact with no more than two data points. The essence of these tasks is to either determine the location of a known event or ascertain the event based on its location. 
In contrast, complex tasks delve into a deeper spatial analysis, encapsulating Associate, Categorize, Cluster, and Rank tasks. 
Here, the exploration revolves around discerning between numerous data points. 
Users predominantly seek data similarities, often independent of specific locations or events.

The comprehensive list of these tasks is presented in Table \ref{tab:exploratory-tasks}. 
Please refer to the supplementary materials for a complete list of questions.
\HDY{
The question lists for the single-map and multiple-map conditions are similar but not identical, as a single-map is effective for gaining insights into a specific geospatial event, whereas multiple-maps are more beneficial for analyzing spatiotemporal discrepancies. 
We are interested in how interactive features have distinct effects on the single-map and multiple-map conditions separately. }
While we acknowledge that the analytical tasks in Table \ref{tab:exploratory-tasks} may span multiple types, we designated each based on its most dominant analytical characteristic. 
These tasks involve multiple-choice questions, where participants select one correct answer from three options.

\subsubsection{Others}
\hspace{\parindent}\textbf{Map Data Context:}
To examine the general effect of interactive features, two datasets with different spatial ranges are utilized in this study (i.e., NY Temperature and CONUS Precipitation).
On one hand, NY Temperature provides temperature data (D\textsubscript{T}) for a limited spatial extent around New York. 
On the other hand, CONUS Precipitation provides precipitation data (D\textsubscript{P}) for the contiguous U.S., covering a larger area than \textit{NY Temperature}.
The numbers following D\textsubscript{T} and D\textsubscript{P} in Table~\ref{tab:task-distribution} refer to the distinct subsets of data. The seven questions for the multiple-map task sets in Table \ref{tab:exploratory-tasks} are developed using the two datasets separately.

\textbf{Participants' Expertise:} We categorize the participants into two groups - novice and experienced, based on their reported expertise in geospatial analysis. 
The participants who identified themselves as \textit{`unfamiliar'} or \textit{`somewhat familiar'} with geospatial analysis are classified as novice users, and the other participants who identified themselves as \textit{`somewhat experienced'}, \textit{`experienced'}, or \textit{`expert'} are classified as experienced users. 
The sample ratio between these two groups was 1.06 which is fairly equal, as Fig. \ref{fig:demographic-data} suggests.

\begin{table}[t]
\centering
\caption{Distribution of the eight sets of tasks that account for the repetitive measures such as the spatial extent, visualization methods, and map layout. Each task set includes a task from the classification of \HDY{the analysis tasks.}
}
\scriptsize
\newcolumntype{P}[1]{>{\centering\arraybackslash}p{#1}}
\begin{tabular} {P{2.5cm} | P{2.6cm} | P{1.1cm}| P{1.1cm}}
\hline
\textbf{Map Data Context} & \multicolumn{1}{c|}{\textbf{\begin{tabular}[c]{@{}c@{}}Map Data\\ (Task Set)\end{tabular}}} & \textbf{\begin{tabular}[c]{@{}c@{}}Participant\\ Group 1\end{tabular}}                          & \textbf{\begin{tabular}[c]{@{}c@{}}Participant\\ Group 2\end{tabular}}                          \\ \hline
                                                                                                                              & \cellcolor[HTML]{EFEFEF}single-map - D\textsubscript{T}1                                                                             & \cellcolor[HTML]{EFEFEF}                                                                         &                                                                                                 \\ 
\multirow{-2}{*}{\begin{tabular}[c]{@{}c@{}}Small Spatial Extent\\ (NY Temperature)\end{tabular}}                             & Multiple-Map - {[}D\textsubscript{T}2, D\textsubscript{T}3{]}                                                                    & \cellcolor[HTML]{EFEFEF}                                                                         &                                                                                                 \\ \cline{1-2}
\cellcolor[HTML]{EFEFEF}                                                                                                      & \cellcolor[HTML]{EFEFEF}single-map - D\textsubscript{P}1                                                                             & \cellcolor[HTML]{EFEFEF}                                                                         &                                                                                                 \\ 
\multirow{-2}{*}{\cellcolor[HTML]{EFEFEF}\begin{tabular}[c]{@{}c@{}}Large Spatial Extent\\ (CONUS Precipitation)\end{tabular}} & Multiple-Map - {[}D\textsubscript{P}2, D\textsubscript{P}3{]}                                                                    & \multirow{-4}{*}{\cellcolor[HTML]{EFEFEF}\begin{tabular}[c]{@{}c@{}}Static\\Maps\end{tabular}} & \multirow{-4}{*}{\begin{tabular}[c]{@{}c@{}}Interactive\\Maps\end{tabular}}                    \\ \cline{1-2}
                                                                                                                              & \cellcolor[HTML]{EFEFEF}single-map - D\textsubscript{T}4                                                                             &                                                                                                  & \cellcolor[HTML]{EFEFEF}                                                                        \\ 
\multirow{-2}{*}{\begin{tabular}[c]{@{}c@{}}Small Spatial Extent\\ (NY Temperature)\end{tabular}}                             & Multiple-Map - {[}D\textsubscript{T}5, D\textsubscript{T}6{]}                                                                    &                                                                                                  & \cellcolor[HTML]{EFEFEF}                                                                        \\ \cline{1-2}
\cellcolor[HTML]{EFEFEF}                                                                                                      & \cellcolor[HTML]{EFEFEF}single-map - D\textsubscript{P}4                                                                             &                                                                                                  & \cellcolor[HTML]{EFEFEF}                                                                        \\
\multirow{-2}{*}{\cellcolor[HTML]{EFEFEF}\begin{tabular}[c]{@{}c@{}}Large Spatial Extent\\ (CONUS Precipitation)\end{tabular}} & Multiple-Map - {[}D\textsubscript{P}5, D\textsubscript{P}6{]}                                                                    & \multirow{-4}{*}{\begin{tabular}[c]{@{}c@{}}Interactive\\Maps\end{tabular}}   & \multirow{-4}{*}{\cellcolor[HTML]{EFEFEF}\begin{tabular}[c]{@{}c@{}}Static\\Maps\end{tabular}} \\ \hline
\end{tabular}
\label{tab:task-distribution}
\end{table}

\subsection{Procedure}

Upon accessing the interface, the participants are first presented with an overview of the study, detailing the procedure, assignments, and the type of interaction data to be captured, culminating in a request for their consent. 
Subsequently, the interface collects basic demographic information from the participants such as level of education, gender, expertise with geospatial data analysis, familiarity with data visualization, and daily computer usage. 
The participants are then transitioned to a training session that introduces the interactive features for the purpose of visual exploration. 
Across its 12 sections, the training modules elucidate the nuances of the contour map, the geospatial data, and the available interactive elements. 
To ensure comprehension, the participants encounter sample questions post each interactive feature's detailed explanation. 
Incorrect answers are met with on-screen guidance leading them to the correct response.
This feedback mechanism is unique to the training phase and is deactivated during the main session. 
Upon training completion, they are seamlessly moved to the main study session.

In the main study session, the participants encounter eight task sets (i.e., the second column in Table \ref{tab:task-distribution}), split into two distinct sections. 
Each set corresponds to a map layout showcasing different geospatial data. 
Within each set, there is a question representing each of the seven task types, culminating in a total of 56 questions (eight task sets $\times$ seven task questions) throughout the study. 
Table \ref{tab:task-distribution} provides a fair distribution of the map and task types, ensuring samples are counterbalanced and repetitive measures are taken into account. 
The tasks are multiple choice type questions regarding the questions in Table~\ref{tab:exploratory-tasks}, and the participants can submit their answers by selecting from the options provided for each task. 
As participants perform the tasks, either by interacting or observing the map visualization, the interface logs their observations, task completion time (in seconds), and usage of any interactive features associated with questions. Upon completing the study, 
They get a completion code from the interface. They then submit the code to the MTurk portal to claim their incentive.

\subsection{Evaluation Metrics}
Evaluating geospatial visualizations involves both qualitative and quantitative metrics \cite{tominski2011information}. 
In our study, we focus on quantitative metrics to evaluate the efficacy of interactive features embedded within the contour map. 
Roth \cite{roth2013empirically} described the objectives of cartographic interaction, to name a few, as (1) presenting insights preemptively, (2) coming to a decision point faster, meaning not solely doing it with a visual inspection. 
To assess the effectiveness of the interactive features \HDY{in analysis tasks} and validate our hypotheses, we focus on the quantifiable factors - \textit{\textbf{the task accuracy and \textit{\textbf{completion time} of their \HDY{performed task}}} -} and \textit{\textbf{interaction logs during the task}}. 
The accuracy measures the participant's analytical ability to correctly answer the given multiple-choice questions. 
\HDY{Each study condition includes two questions and task accuracy is calculated based on the number of correct answers. While the task accuracy is inherently a continuous value, only three discrete values of accuracy are available (0 (0\%), 1 (50\%), and 2 (100\%)).
We consider that the intervals between these scores (0 to 50 and 50 to 100) reflect equal changes in task performance. 
Despite the discrete nature of these values, we treat accuracy as a continuous variable~\cite{robitzsch2020ordinal}, given that the scores are proportional and reflect distinct performance levels on an underlying continuous scale of task accuracy. Therefore, we use continuous analysis methods, as outlined in Section ~\ref{section-result}.
}

\subsection{Hypotheses} \label{sec:hypothesis-rq}

Acknowledging the potential of interactive contours in a geospatial view for visually analyzing geospatial data, it is essential to understand the utility of interactive features to the full extent. 
To address our RQs in Section 1, we establish the following hypotheses for our study:

    \begin{enumerate}[label=\textbf{H\arabic*:}]
    \item Interactive features can improve users' accuracy when \HDY{performing geospatial} analysis tasks. 
    
    \item When the interactive features are available, users might need more time to complete tasks \HDY{compared to using} the static map due to associated cognitive processing.


\end{enumerate}

\section{\HDY{Results}}\label{section-result}

\HDY{We leverage one-way repeated measures analysis of variance (ANOVA) at a 95\% confidence level. 
It evaluates performance metrics based on the map types (interactive vs. static) across the map layouts and task types.} 
Although our results violate the normality test, we use ANOVA because the previous study~\cite{blanca2017non} provided empirical evidence for its robustness when evaluating data with non-normal distributions. To present the estimates, we also report the lower and upper bounds at a 95\% confidence interval (CI). To present differences between visualization methods, we report the $p$-value and the error bound mean ($EBM$) for the task types. 

\subsection{Accuracy}
Fig. \ref{fig:performance-ci} shows the accuracy results by task types, map types, and layouts. 
\subsubsection{\HDY{Multi-Map Result Analysis}}

\hspace{\parindent}\textbf{Categorize:} 
We found a significant main effect of map type ($p=.035$, $F(1, 61)=4.675$, $\eta^2_p=.071$). The accuracy when using the interactive map ($M = .645$, $EBM = [.561, .729]$) was statistically higher than when using the static map ($M = .516$, $EBM = [.427, .605]$) for categorize-type tasks.

\textbf{Cluster:} A significant map type ($p=.014$, $F(1, 61)=6.437$, $\eta^2_p=.095$) showed a main effect. Regarding the map type, its pairwise comparison result showed that the interactive map ($M=.597$, $EBM=[.508, .685]$) supported the participants better than the static map ($M=.444$, $EBM= [.354, .533]$). 

\textbf{Distinguish:} We found a significant main effect of map type ($p=.002$, $F(1, 61)=10.738$, $\eta^2_p=.150$). The pairwise comparison showed the interactive-map condition ($M=.629$, $EBM=[.553, .705]$) showed higher accuracy than the single-map condition ($M=.444$, $EBM=[.354, .533]$).


\begin{figure}[t]
    \centering
    \includegraphics[width=\linewidth]{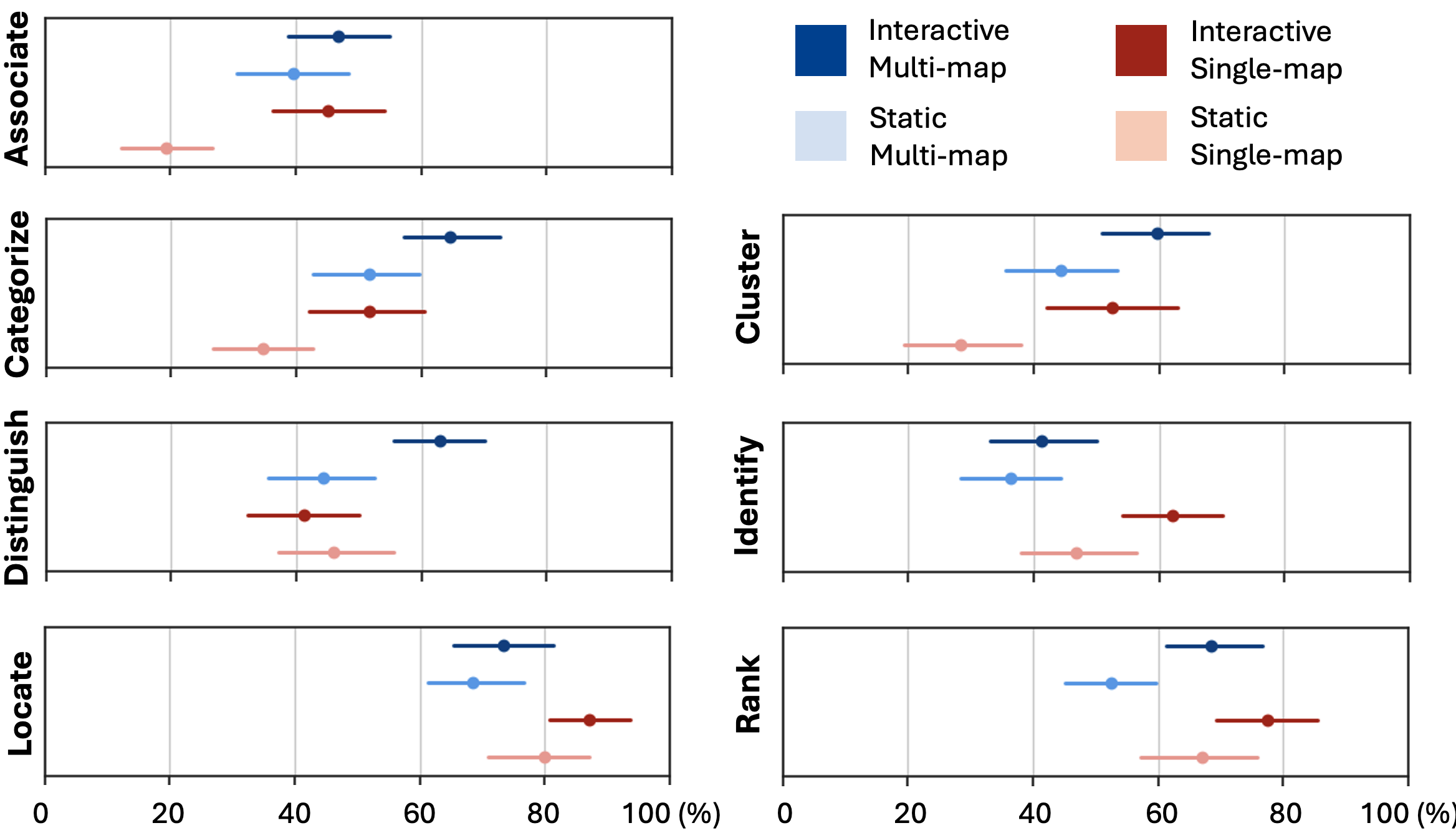}
    \caption{
    The accuracy results by tasks, map types, and map layouts 
    }
    \label{fig:performance-ci}
\end{figure}

\textbf{Rank:} We discovered a significant main effect  ($p=.003$, $F(1, 61)=9.472$, $\eta^2_p=.134$).
The interactive map ($M=.685$, $EBM=[.605, .766]$) had a higher accuracy than the static map ($M=.524$, $EBM=[.443, .605]$).

\textbf{Associate, Identify, \& Locate:} No clear difference in task accuracy was disclosed by the task type.

\subsubsection{\HDY{Single-Map Result Analysis}}
\hspace{\parindent}\textbf{Associate:} We found a significant main effect ($p<.001$, $F(1, 61)=15.870$, $\eta^2_p=.206$) was disclosed. The results suggested that the participants performed substantially better using the interactive map ($M=.452$, $EBM=[.356, .546]$) compared to the static map ($M=.194$, $EBM=[.120, .268]$). 

\textbf{Categorize:} We found a significant main effect of map type ($p=.009$, $F(1, 61)=7.245$, $\eta^2_p=.106$). The accuracy when using the interactive map ($M = .516$, $EBM = [.419, .614]$) was statistically higher than when using the static map ($M = .347$, $EBM = [.262, .432]$).

\textbf{Cluster:} The map type ($p<.001$, $F(1, 61)=14.886$, $\eta^2_p=.196$) showed a significant main effect. Its pairwise comparison result showed that the interactive map ($M=.524$, $EBM=[.415, .633]$) supported the participants better than the static map ($M=.282$, $EBM= [.183, .381]$). 


\textbf{Identify:} A main effect of map type ($p=.029$, $F(1, 61)=4.990$, $\eta^2_p=.076$) was disclosed. The posthoc test revealed \HDY{that} the participants achieved better results with the interactive map ($M=.621$, $EBM=[.533, .709]$) compared to the static map ($M=.468$, $EBM=[.368, .568]$). 



\textbf{Distinguish, Locate, \& Rank:} We found no main effect of task type on the accuracy.

\subsection{Task Completion Time}
Fig. \ref{fig:single-multiple-performance-ci} shows the task completion time results by task types, map types, and layouts. 

\subsubsection{\HDY{Multi-Map Result Analysis}}

\hspace{\parindent}\textbf{Associate:} We found a main effect of map type ($p=.017$, $F(1, 61)=6.042$, $\eta^2_p=.090$). 
The pairwise comparison results showed the participants spent statistically more time on the interactive map ($M=72.711s$, $EBM=[51.814, 93.608]$) than the static map ($M=47.640s$, $EBM=[32.519, 62.761]$). 

\textbf{Categorize:} We found a main effect of map type ($p=.017$, $F(1, 61)=5.996$, $\eta^2_p=.089$). Its pairwise comparison disclosed that the interactive map condition ($M=106.916s$, $EBM=[79.213, 134.619]$) took more time than the static map condition ($M=73.820s$, $EBM=[53.819, 93.822]$).



\textbf{Cluster, Distinguish, Identify, Locate, \& Rank:} No clear difference in task completion time was disclosed by the task type.

\subsubsection{\HDY{Single-Map Result Analysis}}

\hspace{\parindent}\textbf{Cluster:} A main effect of map type ($p=.045$, $F(1, 61)=4.207$, $\eta^2_p=.065$) was reported. Its pairwise comparison disclosed that the interactive map condition ($M=72.304s$, $EBM=[45.746, 98.861]$) took more time than the static map condition ($M=44.161s$, $EBM=[32.593, 55.729]$).

\textbf{Rank:} We found a main effect of map type ($p=.004$, $F(1, 61)=8.899$, $\eta^2_p=.127$). The pairwise comparison results showed the participants spent statistically more time on the interactive map ($M=65.457s$, $EBM=[52.582, 78.332]$) than the static map ($M=47.133s$, $EBM=[39.255, 55.011]$). 


\textbf{Associate, Categorize, Distinguish, Identify, \& Locate:} No clear difference in task completion time was disclosed by the task type and layout.

\begin{figure}[t]
    \centering
    \includegraphics[width=\linewidth]{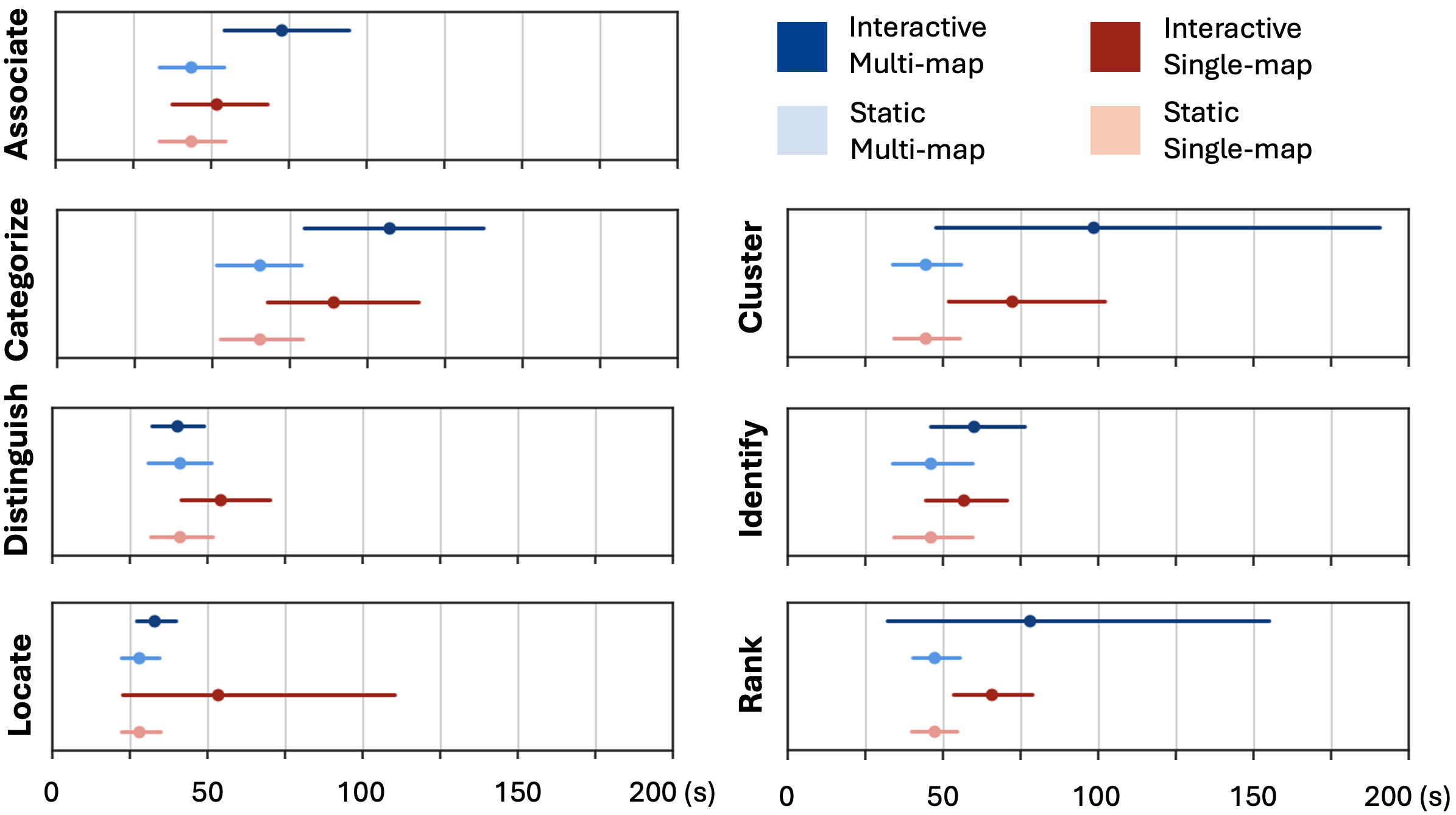}
    \caption{
    The task completion time results by tasks, map types, and map layouts
    }
    \label{fig:single-multiple-performance-ci}
\end{figure}

\section{Discussion}

Our user study results partially support all hypotheses. 
In response to \textbf{RQ1}, we found that the results partially support \textbf{H1}. 
\HDY{On one hand, when the multiple-maps were presented, the interactive features significantly improved accuracy in four out of seven tasks (i.e., Categorize, Cluster, Distinguish, and Rank). On the other hand, in the single-map condition, accuracy was significantly enhanced by the interactive features for four different tasks (i.e., Associate, Categorize, Cluster, and Identify). 
These results align with the perspectives of both DiBiase \cite{dibiase1990visualization} and Roth \cite{roth2008addressing}, indicating that user interaction, particularly in terms of precision performance, aided in uncovering data relationships. However, we found no clear evidence that introducing interactivity improved task accuracy for Identity and Locate tasks in both the single and multiple-map conditions.}

Additionally, the results partially support \textbf{H2}. \HDY{Firstly, despite the performance improvements from interactive features in the multiple-map condition, we observed no significant change in completion time for the Cluster, Distinguish, and Rank tasks. However, the Categorize task took longer to complete. Similarly, in the single-map condition, while performance was enhanced by the interactive features, no noticeable variation in completion time was found for the Associate, Categorize, and Identity tasks, but the Cluster task showed an increased duration.}
In the following section, we explore RQ1 more thoroughly by examining the impact of participant expertise. We then proceed to analyze RQ2, drawing upon the insights gleaned from our study.

\subsection{Observations from the User Interaction Logs}
To better understand participants' usage and pattern of interaction features, we examine interaction logs within the context of \HDY{the analysis task types}.
We classify the interactions based on the number of mouse/keyboard operations (steps) required to perform the feature. Following that, we eliminate unintended interactions in the process of aggregating their interaction logs.
For instance, despite participants only needing to move the cursor to inspect contour regions, they unintentionally interacted with other contour areas while navigating toward deeper parts of the map.
We filter out these unintended interactions by using a time threshold, focusing on areas where participants spent a notable duration, suggesting they were genuinely processing the information displayed on tooltips or widgets.
Fig. \ref{fig:interaction-usage} shows a comprehensive overview of the usage patterns for the interactive features except for synchronization.

\begin{figure}[t]
    \centering
    \includegraphics[width=\linewidth]{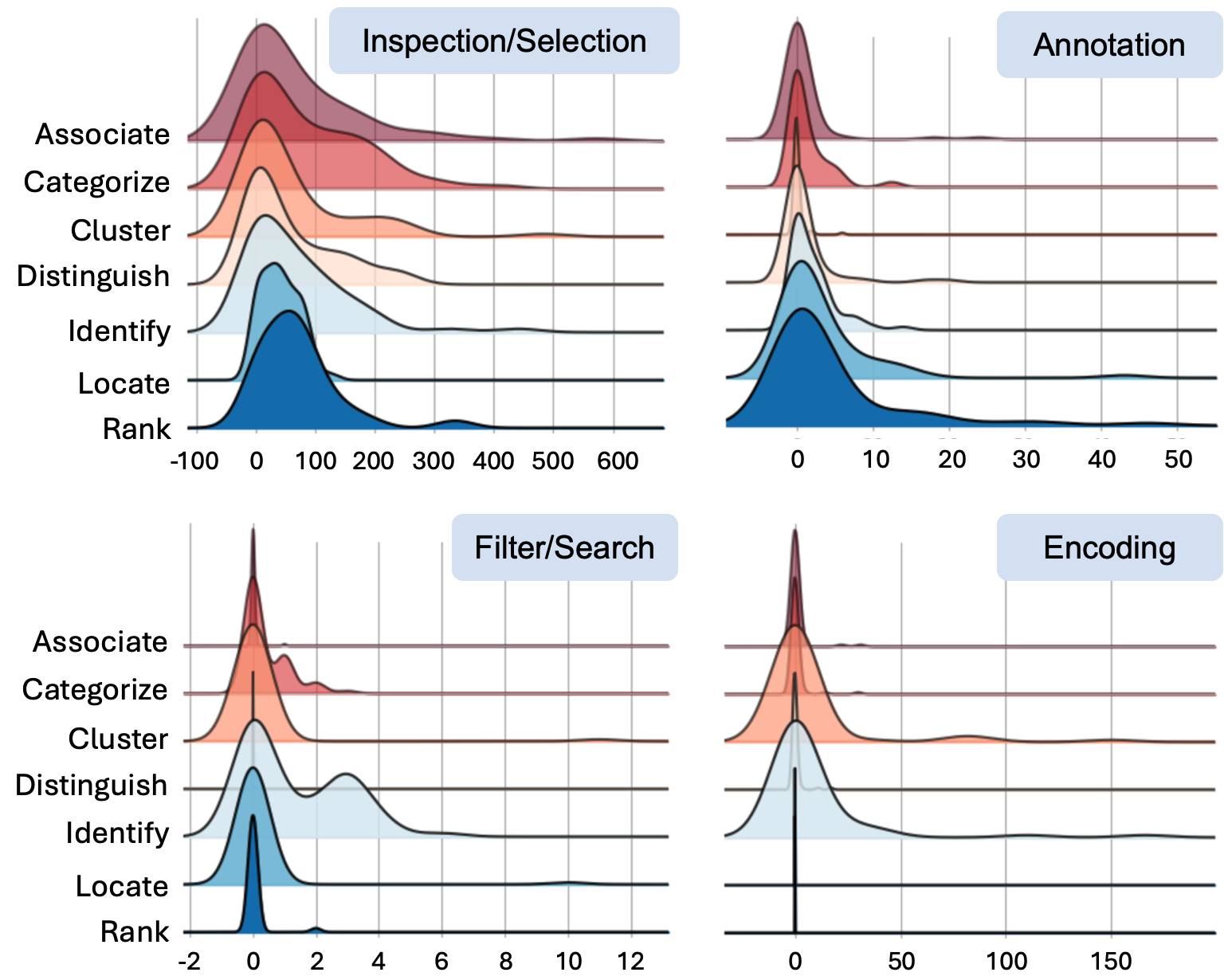}
    \caption{
    The frequently used interactive features in the \HDY{participating tasks.}
    The plotted range of usage frequency is categorized by the task types and bootstrapped with a 95\% confidence interval.
    }
    \label{fig:interaction-usage}
\end{figure}

\textbf{Identifying the frequently used interactive features:} 
We find that inspection and selection are the most frequently used functions, followed by annotation, filter/search, and encoding. 
First of all, we observe that the participants frequently inspected or selected various distinct contour regions during the tasks involving association, categorization, and clustering. We expect that they used these interactive features to understand the relationships among different intensity levels.

Next, annotation is mostly used in the Locate, Rank, and Associate tasks. 
Annotation is particularly useful in the Rank task because the marker widget provides insights for comparison with other annotations, thereby aiding in determining the order of the contour regions. We also observe that the annotation interaction generally comes after the inspection, selection, and filter interactions.

The filter interaction is frequently utilized in the Identify and Categorize tasks, often involving searching based on geolocation. 
The Identify tasks require to extract information based on geographical characteristics. 
Thus, participants used the search feature to annotate or inspect intensities according to the locations. 
The comparative information provided by the annotated locations is continuously updated as the participants interact with other contour regions providing effective categorization and clustering tasks, which involve identifying the regions with similar variability.

At last, we identify that the encoding feature is mostly used in the Cluster and Identify tasks. 
The contour encodings are understandably useful in the Cluster task, as it is often required to identify the unique regions on the contour map. 
The Identify task involves the geographical characteristics that usually reside in the base map, hence, the participants' controlled styles correlate with the contour map in these tasks. Therefore, changing the weight and opacity of the contour lines and region areas can be beneficial in visually performing these tasks.

\begin{figure}[t]
    \centering
    \includegraphics[width=\linewidth]{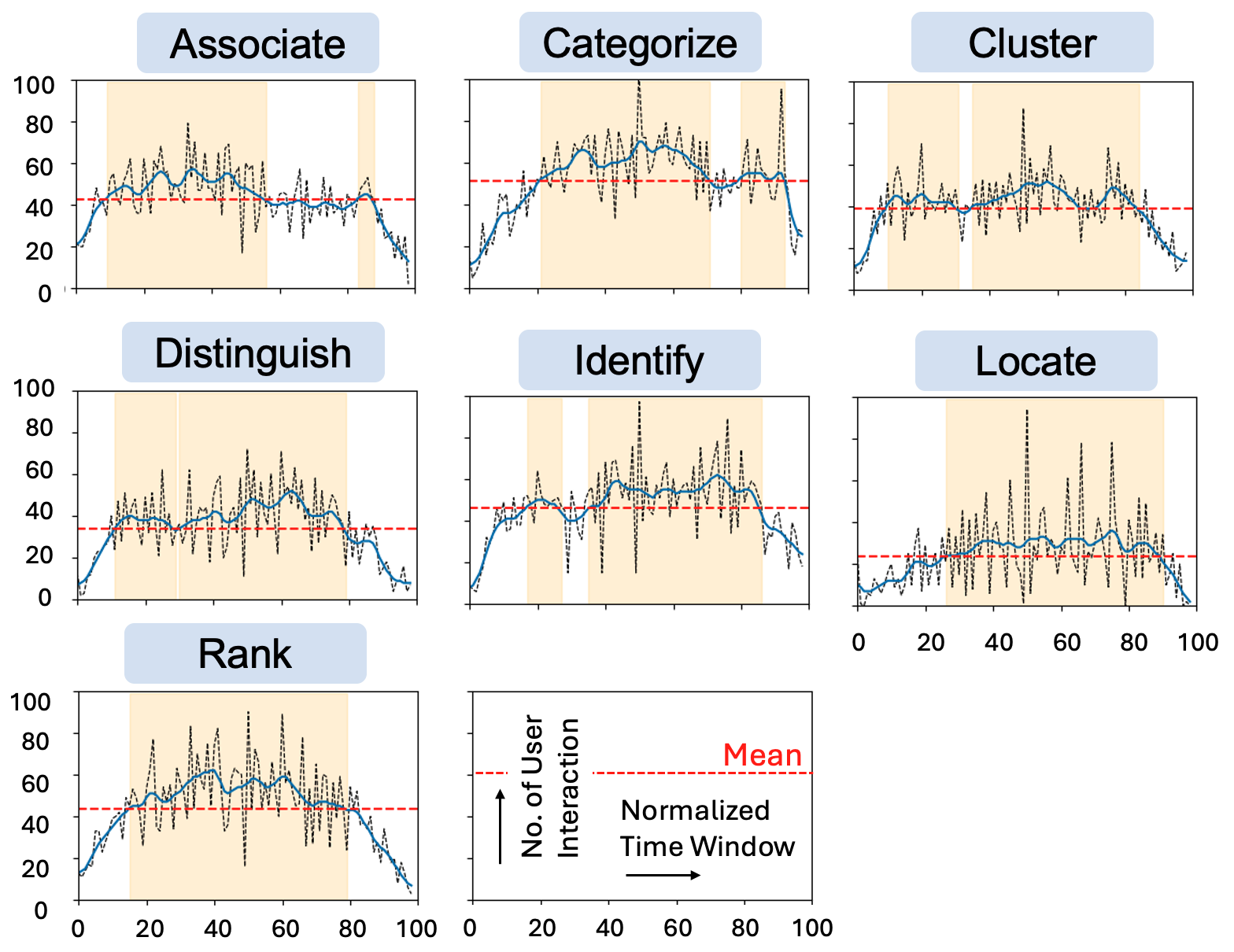}
    \caption{Task-wise user interaction pattern based on the normalized task completion time window. The highlighted area represents the time segments where participants' interactivity exceeds the mean.}
    \label{fig:interaction-pattern}
\end{figure}

\textbf{Reflecting cognitive process with interaction usage:}
Fig. \ref{fig:interaction-pattern} presents the number of user interaction usage over time. We normalize the time based on task completion time for each task.
It provides the opportunity to estimate the segments where the participants are seemingly cognitively active. 
Red lines represent the average number of user interaction usage, and highlighted areas show the windows where participants' interactivity exceeds the average. 
The periods of inactivity, particularly at the beginning and end, reflect participants' cognitive processing as they attempted to comprehend the task at hand, plan their next move, or reach an analytical conclusion.

The average number of user interaction usage by the tasks reveals that the participants were especially interactive during more complex tasks such as associating, categorizing, and ranking. 
They also devoted significant time to cognitively processing relationships between spatial data points.
In contrast, for tasks such as distinguishing and locating, participants were notably less interactive. 
This pattern is further reflected in the accuracy score, where the performance disparity between static and interactive maps is minimal for these specific tasks.

\subsection{Performance Elevation by Geospatial Expertise}

\begin{figure*}[t]
    \centering
    \includegraphics[width=\linewidth]{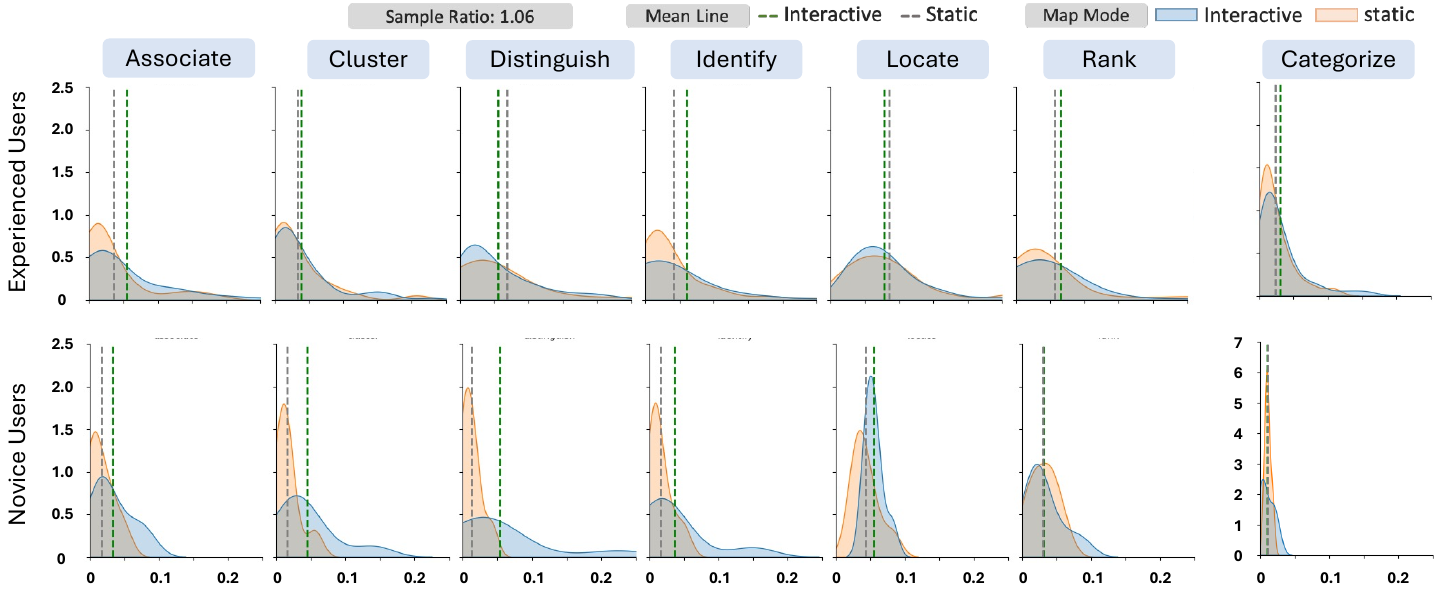}
    \caption{Comparison between novice and experienced participants' performance suggests that novice users excel with interactive features significantly more than experienced users.
    \HDY{The X-axis represents user performance, measured as the total number of correct answers divided by the total task completion time.}
    }
    \label{fig:user-performance-by-expertise}
\end{figure*}

Our results show that the interactive map could improve the novice users' performance when compared to the static map, closing performance gaps between novice and experienced.
This is because interpreting information on a static map typically demands a certain level of expertise, whereas interactive maps have the potential to facilitate a better understanding for both expert and novice users alike. 
We examine the participants' task accuracy per unit of time across the \HDY{seven analysis} task types. 
It is calculated by \HDY{dividing} the total number of correct answers by the total amount of task completion time. 
Fig. \ref{fig:user-performance-by-expertise} illustrates task-wise density plots to separately compare the performance measures of both novice and experienced users in interactive and static maps respectively. 
The dashed lines in Fig. \ref{fig:user-performance-by-expertise} represent the performance mean for static (gray) and interactive (green) maps. 
It shows that experienced users performed significantly better than novice users. 
It also shows that the interactive features improved novice users' performance in the associate, cluster, and rank tasks more than experienced users.
At the same time, however, the results reveal that the Categorize task is found to be particularly difficult for novice users compared to the other tasks.

 \subsection{Design Considerations for Interactive Contour Map\HDY{s}}
\begin{itemize}
    \item \textbf{Prioritize Interactive Features:} Emphasize the development and improvement of interactive features.
    Such features can significantly influence users' analytical performance.

    \item \textbf{Consider User Expertise:} It is important to understand how performance varies with the expertise levels of end-users. Interactive maps can be particularly helpful for novice users, effectively bridging the performance gaps seen with traditional static maps.

    \item \textbf{Address Interaction Efficiency:} It is crucial to note that while interactive maps may improve accuracy, they might also increase task completion time. It is important to refine the efficiency of interactive features to ensure a balance between accuracy and time efficiency.

    \item \textbf{Tailor Interactions to Task Types:} Interactive features should be customized based on \HDY{the analysis task types}. 
    For instance, emphasize inspection and selection for single-step interactions and consider multi-step interactions for tasks such as locating.
    In addition, the value of annotations, especially in tasks involving ranking and associating cannot be overstated. Users should be encouraged to engage with annotations after inspection, selection, and filter interactions to help their decision-making process.

\end{itemize}

\subsection{Future Work}

In this work, we found that interactive features could significantly improve correctness in geospatial data analysis tasks. However, we observed that the participants struggled in \HDY{complex analysis tasks} such as the Associate, Categorize, and Cluster tasks which require in-depth spatial correlation and comparison compared to the simple Identify and Locate tasks. 
Additionally, the overall correctness for the tasks is less than 70\%, indicating that the navigation task accuracy still needs to be improved.
Interactive features can effectively support users in data exploration beyond individual visualization elements by facilitating coordinated interactions among different visual components.
In our future research, we aim to investigate visualization components which could support the interactive map view and the benefits of coordinated interaction between them. 
It also involves evaluating visualization approaches with various domain-specific experts in geographic spatial analysis, identifying their key tasks within each domain, and evaluating interactive features that effectively support their tasks in terms of both quantitative and qualitative assessments.
\section{Conclusion}
In this work, we examined the usability of contour-based interactive geospatial visualizations \HDY{in analysis tasks.}
We implemented and designed \HDY{a} web interface and conducted \HDY{a} crowd-sourced user study to quantitatively evaluate and compare user performance when using interactive and static geospatial maps. 
\HDY{The results suggested that using interactive maps improved task accuracy but increased completion time compared to using the static maps.}
We also analyzed participants' interaction logs to discuss the usability of the interactive features in \HDY{the geospatial visualization tasks.} 
Our findings can be used as design considerations for future visualization researchers and practitioners to create use case-specific interactive geovisualizations for exploring geospatial data.

\section*{Acknowledgments}
This paper was supported by the US National Science Foundation (NSF) Data Infrastructure Building Blocks
(DIBBs) Program (Award \#1640818).
\bibliographystyle{IEEEtran}
\bibliography{0_reference}

\begin{thebibliography}{10}
\providecommand{\url}[1]{#1}
\csname url@samestyle\endcsname
\providecommand{\newblock}{\relax}
\providecommand{\bibinfo}[2]{#2}
\providecommand{\BIBentrySTDinterwordspacing}{\spaceskip=0pt\relax}
\providecommand{\BIBentryALTinterwordstretchfactor}{4}
\providecommand{\BIBentryALTinterwordspacing}{\spaceskip=\fontdimen2\font plus
\BIBentryALTinterwordstretchfactor\fontdimen3\font minus \fontdimen4\font\relax}
\providecommand{\BIBforeignlanguage}[2]{{%
\expandafter\ifx\csname l@#1\endcsname\relax
\typeout{** WARNING: IEEEtran.bst: No hyphenation pattern has been}%
\typeout{** loaded for the language `#1'. Using the pattern for}%
\typeout{** the default language instead.}%
\else
\language=\csname l@#1\endcsname
\fi
#2}}
\providecommand{\BIBdecl}{\relax}
\BIBdecl

\bibitem{McLean2020}
``The hawai`i rainfall analysis and mapping application (hi-rama): Decision support and data visualization for statewide rainfall data.''\hskip 1em plus 0.5em minus 0.4em\relax Association for Computing Machinery, 7 2020, pp. 239--245.

\bibitem{Alder2015}
J.~R. Alder and S.~W. Hostetler, ``Web based visualization of large climate data sets,'' \emph{Environmental Modelling \& Software}, vol.~68, pp. 175--180, 2015.

\bibitem{li2020sovas}
Z.~Li, Q.~Huang, Y.~Jiang, and F.~Hu, ``Sovas: a scalable online visual analytic system for big climate data analysis,'' \emph{International Journal of Geographical Information Science}, vol.~34, no.~6, pp. 1188--1209, 2020.

\bibitem{wang2014}
Y.~Q. Wang, ``Meteoinfo: Gis software for meteorological data visualization and analysis,'' \emph{Meteorological Applications}, vol.~21, no.~2, pp. 360--368, 2014.

\bibitem{Sharma2018}
A.~Sharma, S.~M.~A. Zaidi, V.~Chandola, M.~R. Allen, and B.~L. Bhaduri, ``Webglobe - a cloud-based geospatial analysis framework for interacting with climate data.''\hskip 1em plus 0.5em minus 0.4em\relax Association for Computing Machinery, Inc, 11 2018, pp. 42--46.

\bibitem{wang2019open}
Y.~Wang, ``An open source software suite for multi-dimensional meteorological data computation and visualisation,'' \emph{Journal of Open Research Software}, vol.~7, 07 2019.

\bibitem{8613864}
F.~Kamw, S.~Al-Dohuki, Y.~Zhao, T.~Eynon, D.~Sheets, J.~Yang, X.~Ye, and W.~Chen, ``Urban structure accessibility modeling and visualization for joint spatiotemporal constraints,'' \emph{IEEE Transactions on Intelligent Transportation Systems}, vol.~21, no.~1, pp. 104--116, 2020.

\bibitem{zeng2014visualizing}
W.~Zeng, C.-W. Fu, S.~M. Arisona, A.~Erath, and H.~Qu, ``Visualizing mobility of public transportation system,'' \emph{IEEE transactions on visualization and computer graphics}, vol.~20, no.~12, pp. 1833--1842, 2014.

\bibitem{feng2020topology}
Z.~Feng, H.~Li, W.~Zeng, S.-H. Yang, and H.~Qu, ``Topology density map for urban data visualization and analysis,'' \emph{IEEE transactions on visualization and computer graphics}, vol.~27, no.~2, pp. 828--838, 2020.

\bibitem{Steed2020}
C.~A. Steed, J.~R. Goodall, J.~Chae, and A.~Trofimov, ``Crossvis: A visual analytics system for exploring heterogeneous multivariate data with applications to materials and climate sciences,'' \emph{Graphics and Visual Computing}, vol.~3, p. 200013, 2020.

\bibitem{nayeem2021visual}
A.-A.-R. Nayeem, M.~Elshambakey, T.~Dobbs, H.~Lee, D.~Crichton, Y.~Zhu, C.~Chokwitthaya, W.~J. Tolone, and I.~Cho, ``A visual analytics framework for distributed data analysis systems,'' in \emph{2021 IEEE International Conference on Big Data (Big Data)}, Dec 2021, pp. 229--240.

\bibitem{maceachren1997exploratory}
A.~M. MacEachren and M.-J. Kraak, ``Exploratory cartographic visualization: advancing the agenda,'' pp. 335--343, 1997.

\bibitem{andrienko2007geo}
G.~Andrienko, N.~Andrienko, P.~Jankowski, D.~Keim, M.-J. Kraak, A.~MacEachren, and S.~Wrobel, ``Geovisual analytics for spatial decision support: Setting the research agenda,'' \emph{International journal of geographical information science}, vol.~21, no.~8, pp. 839--857, 2007.

\bibitem{robinson2017}
A.~C. Robinson, D.~J. Peuquet, S.~Pezanowski, F.~A. Hardisty, and B.~Swedberg, ``Design and evaluation of a geovisual analytics system for uncovering patterns in spatio-temporal event data,'' \emph{Cartography and Geographic Information Science}, vol.~44, no.~3, pp. 216--228, 2017.

\bibitem{andrienko2001choropleth}
G.~Andrienko, N.~Andrienko, and A.~Savinov, ``Choropleth maps: classification revisited,'' in \emph{Proceedings ica}, 2001, pp. 1209--1219.

\bibitem{nusrat2016state}
S.~Nusrat and S.~Kobourov, ``The state of the art in cartograms,'' in \emph{Computer Graphics Forum}, vol.~35, no.~3.\hskip 1em plus 0.5em minus 0.4em\relax Wiley Online Library, 2016, pp. 619--642.

\bibitem{tominski2011information}
C.~Tominski, J.~F. Donges, and T.~Nocke, ``Information visualization in climate research,'' in \emph{2011 15th International Conference on Information Visualisation}.\hskip 1em plus 0.5em minus 0.4em\relax IEEE, 2011, pp. 298--305.

\bibitem{afzal2019state}
S.~Afzal, M.~M. Hittawe, S.~Ghani, T.~Jamil, O.~Knio, M.~Hadwiger, and I.~Hoteit, ``The state of the art in visual analysis approaches for ocean and atmospheric datasets,'' in \emph{Computer Graphics Forum}, vol.~38, no.~3.\hskip 1em plus 0.5em minus 0.4em\relax Wiley Online Library, 2019, pp. 881--907.

\bibitem{CHEN2019129}
X.~Chen, L.~Shen, Z.~Sha, R.~Liu, S.~Chen, G.~Ji, and C.~Tan, ``A survey of multi-space techniques in spatio-temporal simulation data visualization,'' \emph{Visual Informatics}, vol.~3, no.~3, pp. 129--139, 2019.

\bibitem{andrienko2020spatio}
N.~Andrienko and G.~Andrienko, ``Spatio-temporal visual analytics: a vision for 2020s,'' \emph{Journal of Spatial Information Science}, no.~20, pp. 87--95, 2020.

\bibitem{nayeem2022dcpviz}
A.-A.-R. Nayeem, H.~Lee, D.~Han, M.~Elshambakey, W.~J. Tolone, T.~Dobbs, D.~Crichton, and I.~Cho, ``Dcpviz: A visual analytics approach for downscaled climate projections,'' in \emph{2022 IEEE International Conference on Big Data (Big Data)}, 2022, pp. 291--300.

\bibitem{gdlviz2022}
A.-A.-R. Nayeem, I.~Segovia-Dominguez, H.~Lee, D.~Han, Y.~Chen, Z.~Zhen, Y.~Gel, and I.~Cho, ``Learning on health fairness and environmental justice via interactive visualization,'' in \emph{2022 IEEE International Conference on Big Data (Big Data)}, 2022, pp. 784--791.

\bibitem{gibson2017use}
P.~B. Gibson, S.~E. Perkins-Kirkpatrick, P.~Uotila, A.~S. Pepler, and L.~V. Alexander, ``On the use of self-organizing maps for studying climate extremes,'' \emph{Journal of Geophysical Research: Atmospheres}, vol. 122, no.~7, pp. 3891--3903, 2017.

\bibitem{Lee2014}
J.~W. Lee and S.~Y. Hong, ``Potential for added value to downscaled climate extremes over korea by increased resolution of a regional climate model,'' \emph{Theoretical and Applied Climatology}, vol. 117, no. 3-4, pp. 667--677, 2014.

\bibitem{Lee2017}
H.~Lee, D.~E. Waliser, R.~Ferraro, T.~Iguchi, C.~D. Peters-Lidard, B.~J. Tian, P.~C. Loikith, and D.~B. Wright, ``Evaluating hourly rainfall characteristics over the us great plains in dynamically downscaled climate model simulations using nasa-unified wrf,'' \emph{Journal of Geophysical Research-Atmospheres}, vol. 122, no.~14, pp. 7371--7384, 2017.

\bibitem{DeSales2013}
F.~De~Sales and Y.~K. Xue, ``Dynamic downscaling of 22-year cfs winter seasonal hindcasts with the ucla-eta regional climate model over the united states,'' \emph{Climate Dynamics}, vol.~41, no.~2, pp. 255--275, 2013.

\bibitem{andrews2010space}
C.~Andrews, A.~Endert, and C.~North, ``Space to think: large high-resolution displays for sensemaking,'' in \emph{Proceedings of the Conference on Human Factors in Computing Systems (CHI)}.\hskip 1em plus 0.5em minus 0.4em\relax ACM, 2010, pp. 55--64.

\bibitem{bradel2013large}
L.~Bradel, A.~Endert, K.~Koch, C.~Andrews, and C.~North, ``Large high resolution displays for co-located collaborative sensemaking: Display usage and territoriality,'' \emph{International Journal of Human-Computer Studies}, vol.~71, no.~11, pp. 1078--1088, 2013.

\bibitem{leigh2019usage}
J.~Leigh, D.~Kobayashi, N.~Kirshenbaum, T.~Wooton, A.~Gonzalez, L.~Renambot, A.~Johnson, M.~Brown, A.~Burks, K.~Bharadwaj \emph{et~al.}, ``Usage patterns of wideband display environments in e-science research, development and training,'' in \emph{2019 15th International Conference on eScience (eScience)}.\hskip 1em plus 0.5em minus 0.4em\relax IEEE, 2019, pp. 301--310.

\bibitem{OGAO200223}
P.~Ogao and M.-J. Kraak, ``Defining visualization operations for temporal cartographic animation design,'' \emph{International Journal of Applied Earth Observation and Geoinformation}, vol.~4, no.~1, pp. 23--31, 2002.

\bibitem{koua2006}
E.~L. Koua, A.~MacEachren, and M.-J. Kraak, ``Evaluating the usability of visualization methods in an exploratory geovisualization environment,'' \emph{International Journal of Geographical Information Science}, vol.~20, no.~4, pp. 425--448, 2006.

\bibitem{poweroflocation}
Fobes, ``The power of place,'' \url{https://i.forbesimg.com/forbesinsights/pitney_bowes_power_of_place/PowerOfPlace.pdf}, accessed May 18, 2024.

\bibitem{Maskey2020}
M.~Maskey, R.~Ramachandran, I.~Gurung, M.~Ramasubramanian, B.~Freitag, A.~Kaulfus, G.~Priftis, D.~Bollinger, R.~Mestre, and D.~da~Silva, ``Employing deep learning to enable visual exploration of earth science events,'' in \emph{IGARSS 2020 - 2020 IEEE International Geoscience and Remote Sensing Symposium}, Sep. 2020, pp. 2248--2251.

\bibitem{Tominski2021}
C.~Tominski, G.~Andrienko, N.~Andrienko, S.~Bleisch, S.~I. Fabrikant, E.~Mayr, S.~Miksch, M.~Pohl, and A.~Skupin, ``Toward flexible visual analytics augmented through smooth display transitions,'' \emph{Visual Informatics}, vol.~5, pp. 28--38, 9 2021.

\bibitem{Quinan2016}
P.~S. Quinan and M.~Meyer, ``Visually comparing weather features in forecasts,'' \emph{IEEE Transactions on Visualization and Computer Graphics}, vol.~22, no.~1, pp. 389--398, Jan 2016.

\bibitem{2018AGUFMIN21B35A}
J.~G. {Acker}, G.~T. {Alcott}, M.~{Ventura}, J.~C. {Wei}, and D.~J. {Meyer}, ``{The Advantages of Synergy - Quantitative Earth Science Data Visualization and Analysis with Giovanni, Panoply, and Excel},'' in \emph{AGU Fall Meeting Abstracts}, vol. 2018, Dec. 2018, pp. IN21B--35.

\bibitem{Hugentobler2008}
M.~Hugentobler, \emph{Quantum GIS}.\hskip 1em plus 0.5em minus 0.4em\relax Boston, MA: Springer US, 2008, pp. 935--939.

\bibitem{hanselman1996mastering}
D.~Hanselman and B.~Littlefield, \emph{Mastering MATLAB: a comprehensive tutorial and reference}.\hskip 1em plus 0.5em minus 0.4em\relax Prentice-Hall, Inc., 1996.

\bibitem{NCAR2019}
D.~Brown, R.~Brownrigg, M.~Haley, and W.~Huang, ``\BIBforeignlanguage{en}{Ncar command language (ncl)},'' 2012.

\bibitem{Berman2001}
F.~Berman, A.~Chien, K.~Cooper, J.~Dongarra, I.~Foster, D.~Gannon, L.~Johnsson, K.~Kennedy, C.~Kesselman, J.~Mellor-Crumme, D.~Reed, L.~Torczon, and R.~Wolski, ``The grads project: Software support for high-level grid application development,'' \emph{The International Journal of High Performance Computing Applications}, vol.~15, no.~4, pp. 327--344, 2001.

\bibitem{CAO2021145320}
D.~Cao, J.~Zhang, L.~Xun, S.~Yang, J.~Wang, and F.~Yao, ``Spatiotemporal variations of global terrestrial vegetation climate potential productivity under climate change,'' \emph{Science of The Total Environment}, vol. 770, p. 145320, 2021.

\bibitem{williams2002climate}
D.~N. Williams, R.~S. Drach, P.~F. Dubois, C.~Doutriaux, C.~J. O’Connor, K.~M. AchutaRao, and M.~Fiorino, ``Climate data analysis tool: An open software system approach,'' in \emph{13th Symp. on Global Change and Climate Variations}, 2002.

\bibitem{potter2009visualization}
K.~Potter, A.~Wilson, P.-T. Bremer, D.~Williams, C.~Doutriaux, V.~Pascucci, and C.~Johhson, ``Visualization of uncertainty and ensemble data: Exploration of climate modeling and weather forecast data with integrated visus-cdat systems,'' in \emph{Journal of Physics: Conference Series}, vol. 180, no.~1.\hskip 1em plus 0.5em minus 0.4em\relax IOP Publishing, 2009, p. 012089.

\bibitem{MEYER2019166}
D.~Meyer and M.~Riechert, ``Open source qgis toolkit for the advanced research wrf modelling system,'' \emph{Environmental Modelling \& Software}, vol. 112, pp. 166--178, 2019.

\bibitem{Herring2017}
J.~Herring, M.~S. VanDyke, R.~G. Cummins, and F.~Melton, ``Communicating local climate risks online through an interactive data visualization,'' \emph{Environmental Communication}, vol.~11, pp. 90--105, 1 2017.

\bibitem{wang2018}
F.~Wang, W.~Li, S.~Wang, and C.~R. Johnson, ``Association rules-based multivariate analysis and visualization of spatiotemporal climate data,'' \emph{ISPRS International Journal of Geo-Information}, vol.~7, no.~7, 2018.

\bibitem{johansson2010evaluating}
J.~Johansson, T.-S.~S. Neset, and B.-O. Linn{\'e}r, ``Evaluating climate visualization: An information visualization approach,'' in \emph{2010 14th International Conference Information Visualisation}.\hskip 1em plus 0.5em minus 0.4em\relax IEEE, 2010, pp. 156--161.

\bibitem{nagel2012}
T.~Nagel, E.~Duval, and A.~Vande~Moere, ``Interactive exploration of geospatial network visualization,'' in \emph{CHI '12 Extended Abstracts on Human Factors in Computing Systems}, ser. CHI EA '12.\hskip 1em plus 0.5em minus 0.4em\relax New York, NY, USA: Association for Computing Machinery, 2012, p. 557–572.

\bibitem{mahmood2019improving}
T.~Mahmood, W.~Fulmer, N.~Mungoli, J.~Huang, and A.~Lu, ``Improving information sharing and collaborative analysis for remote geospatial visualization using mixed reality,'' in \emph{2019 IEEE International Symposium on Mixed and Augmented Reality (ISMAR)}.\hskip 1em plus 0.5em minus 0.4em\relax IEEE, 2019, pp. 236--247.

\bibitem{herman2018evaluation}
L.~Herman, V.~Ju{\v{r}}{\'\i}k, Z.~Stacho{\v{n}}, D.~Vrb{\'\i}k, J.~Russn{\'a}k, and T.~{\v{R}}ezn{\'\i}k, ``Evaluation of user performance in interactive and static 3d maps,'' \emph{ISPRS International Journal of Geo-Information}, vol.~7, no.~11, p. 415, 2018.

\bibitem{amini2014impact}
F.~Amini, S.~Rufiange, Z.~Hossain, Q.~Ventura, P.~Irani, and M.~J. McGuffin, ``The impact of interactivity on comprehending 2d and 3d visualizations of movement data,'' \emph{IEEE transactions on visualization and computer graphics}, vol.~21, no.~1, pp. 122--135, 2014.

\bibitem{lateh2005study}
H.~H. Lateh and A.~Raman, ``A study on the use of interactive web-based maps in the learning and teaching of geography,'' \emph{Malaysian Online Journal of Instructional Technology}, vol.~2, no.~3, pp. 99--105, 2005.

\bibitem{roth2013empirically}
R.~E. Roth, ``An empirically-derived taxonomy of interaction primitives for interactive cartography and geovisualization,'' \emph{IEEE transactions on visualization and computer graphics}, vol.~19, no.~12, pp. 2356--2365, 2013.

\bibitem{hahmann2015contour}
T.~Hahmann and E.~L. Usery, ``What is in a contour map?'' in \emph{International Conference on Spatial Information Theory}.\hskip 1em plus 0.5em minus 0.4em\relax Springer, 2015, pp. 375--399.

\bibitem{keller1994visual}
P.~R. Keller, M.~M. Keller, S.~Markel, A.~J. Mallinckrodt, and S.~McKay, ``Visual cues: practical data visualization,'' \emph{Computers in Physics}, vol.~8, no.~3, pp. 297--298, 1994.

\bibitem{zahan2021contour}
G.~M.~H. Zahan, D.~Mondal, and C.~Gutwin, ``Contour line stylization to visualize multivariate information,'' in \emph{Graphics Interface 2021}, 2021.

\bibitem{lu2017state}
Y.~Lu, R.~Garcia, B.~Hansen, M.~Gleicher, and R.~Maciejewski, ``The state-of-the-art in predictive visual analytics,'' in \emph{Computer Graphics Forum}, vol.~36, no.~3.\hskip 1em plus 0.5em minus 0.4em\relax Wiley Online Library, 2017, pp. 539--562.

\bibitem{dibiase1990visualization}
D.~DiBiase, ``Visualization in the earth sciences,'' \emph{Earth and Mineral Sciences}, vol.~59, no.~2, pp. 13--18, 1990.

\bibitem{openstreetmap}
OpenStreetMap, ``Openstreetmap,'' \url{https://www.openstreetmap.org/}, accessed May 18, 2024.

\bibitem{nytemperaturedata}
C.~Beale, H.~Norouzi, Z.~Sharifnezhadazizi, A.~R. Bah, P.~Yu, Y.~Yu, R.~Blake, A.~Vaculik, and J.~Gonzalez-Cruz, ``Comparison of diurnal variation of land surface temperature from goes-16 abi and modis instruments,'' \emph{IEEE Geoscience and Remote Sensing Letters}, vol.~17, no.~4, pp. 572--576, 2020.

\bibitem{Nemani2015}
R.~R. Nemani, B.~L. Thrasher, W.~Wang, T.~J. Lee, F.~S. Melton, J.~L. Dungan, and A.~Michaelis, ``{NASA Earth Exchange (NEX) Supporting Analyses for National Climate Assessments},'' in \emph{AGU Fall Meeting Abstracts}, vol. 2015, 2015, pp. GC21E--04.

\bibitem{heer2010}
J.~Heer and M.~Bostock, ``Crowdsourcing graphical perception: Using mechanical turk to assess visualization design,'' in \emph{Proceedings of the SIGCHI Conference on Human Factors in Computing Systems}, ser. CHI '10.\hskip 1em plus 0.5em minus 0.4em\relax New York, NY, USA: Association for Computing Machinery, 2010, p. 203–212.

\bibitem{karduni2021}
A.~Karduni, D.~Markant, R.~Wesslen, and W.~Dou, ``A bayesian cognition approach for belief updating of correlation judgement through uncertainty visualizations,'' \emph{IEEE Transactions on Visualization and Computer Graphics}, vol.~27, no.~2, pp. 978--988, 2021.

\bibitem{robitzsch2020ordinal}
A.~Robitzsch, ``Why ordinal variables can (almost) always be treated as continuous variables: Clarifying assumptions of robust continuous and ordinal factor analysis estimation methods,'' in \emph{Frontiers in education}, vol.~5.\hskip 1em plus 0.5em minus 0.4em\relax Frontiers Media SA, 2020, p. 589965.

\bibitem{blanca2017non}
M.~J. Blanca~Mena, R.~Alarc{\'o}n~Postigo, J.~Arnau~Gras, R.~Bono~Cabr{\'e}, and R.~Bendayan, ``Non-normal data: Is anova still a valid option?'' \emph{Psicothema, 2017, vol. 29, num. 4, p. 552-557}, 2017.

\bibitem{roth2008addressing}
R.~E. Roth and M.~Harrower, ``Addressing map interface usability: Learning from the lakeshore nature preserve interactive map,'' \emph{Cartographic Perspectives}, no.~60, pp. 46--66, 2008.

\end{thebibliography}


\section{Biography Section}
 




\begin{IEEEbiography}
[{\includegraphics[width=1in,height=1.25in,clip,keepaspectratio]{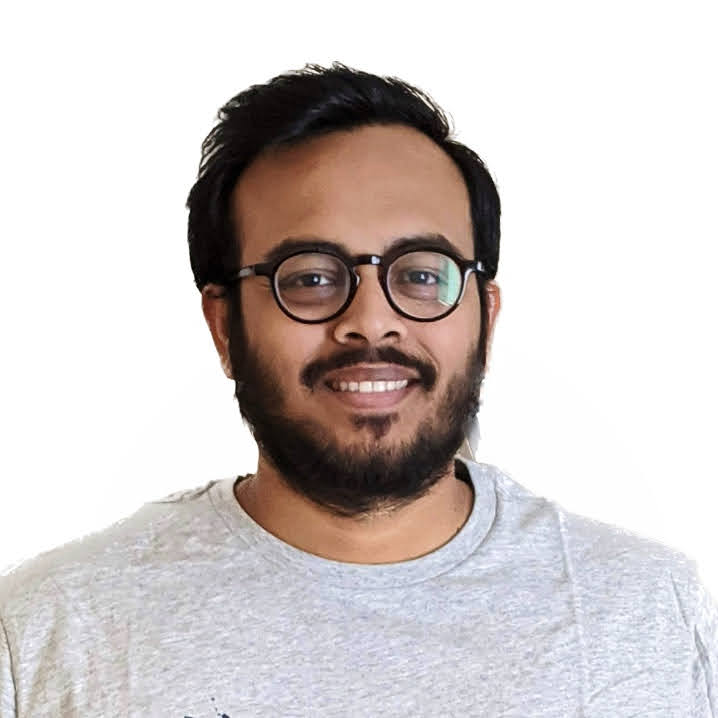}}]
{Abdullah-Al-Raihan Nayeem (co-first author)}
Abdullah-Al-Raihan Nayeem has a Ph.D. in Computing \& Information Systems from the University of North Carolina at Charlotte. He has a robust background in designing exploratory and predictive visual data pipelines. His research interests are visual analytics systems, geo-spatiotemporal visualization, and distributed data analysis. 

\end{IEEEbiography}

\begin{IEEEbiography}[{\includegraphics[width=1in,height=1.25in,clip,keepaspectratio]{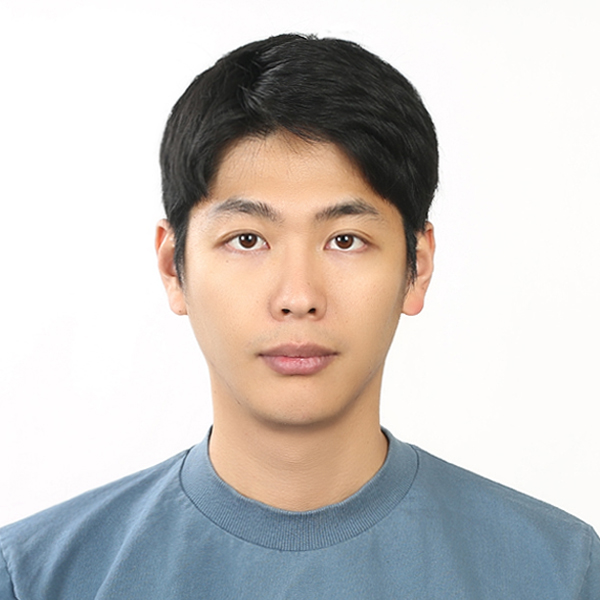}}]{Dongyun Han (co-first author)} 
Dongyun Han is a Ph.D. candidate in the Computer Science Department at Utah State University. His research interests lie in the design, development, and evaluation of visualization and immersive environment systems to help humans communicate with data for sense-making and decision making process.
\end{IEEEbiography}

\begin{IEEEbiography}[{\includegraphics[width=1in,height=1.25in,clip,keepaspectratio]{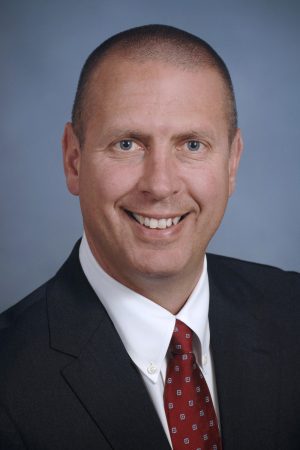}}]{William J. Tolone}
William J. Tolone is a professor and associate dean of the College of Computing and Informatics at the University of North Carolina at Charlotte. He received his Ph.D. in Computer Science from the University of Illinois at Urbana-Champaign. His research expertise is in integrated modeling and simulation, critical infrastructure analytics, visual and data analytics, and collaborative systems. 
\end{IEEEbiography}

\begin{IEEEbiography}[{\includegraphics[width=1in,height=1.25in,clip,keepaspectratio]{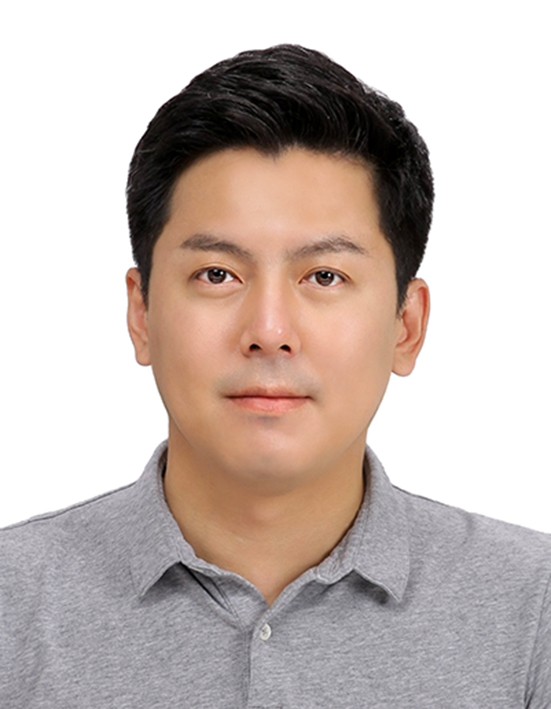}}]{Isaac Cho}
Isaac Cho is an assistant professor in the Computer Science department at Utah State University and an adjunct professor in the Computer Science department at the University of North Carolina at Charlotte. He received his Ph.D. in Computer Science from the University of North Carolina at Charlotte. His main research interests are interactive visual analytics, immersive analytics, data visualization, Mixed Reality, and human-computer interactions. 
\end{IEEEbiography}

\vfill

\end{document}